\documentclass{elsart}
\usepackage{graphicx,epsfig,epsf,amssymb}
\usepackage{amssymb}
\usepackage{longtable}

\newcommand{\dg}{\ensuremath{^\circ}}
\newcommand{\hess}{H.E.S.S.}

\begin{document}

\begin{frontmatter}

\title{Selection and 3D-Reconstruction of Gamma-Ray-induced Air Showers with 
a Stereoscopic System of Atmospheric Cherenkov Telescopes}

\author[LLR]{M. Lemoine-Goumard\corauthref{add1}}
\author[LLR]{B. Degrange}
\author[LLR,LEA]{M. Tluczykont}
\address[LLR]{
Laboratoire Leprince-Ringuet, IN2P3/CNRS,
Ecole~Polytechnique, F-91128~Palaiseau Cedex, France}
\address[LEA]{
 European Associated Laboratory for gamma-ray astronomy, jointly 
supported by CNRS and MPG}

\corauth[add1]{Corresponding author: M. Lemoine-Goumard, e-mail address: lemoine@poly.in2p3.fr}

\begin{abstract}
A simple 3D-reconstruction method for gamma-ray induced air showers is presented,
which takes full advantage of the assets of a system of
Atmospheric Cherenkov Telescopes combining stereoscopy and 
fine-grain imaging like the High Energy Stereoscopic System (\hess). The rich information collected by the cameras allows
to select electromagnetic showers on the basis of their
rotational symmetry with respect to the incident direction, as well as of
their relatively small lateral spread. In the framework of a 3D-model of
the shower, its main parameters --- incident direction, shower core
position on the ground, slant depth of shower maximum, average lateral spread of 
Cherenkov photon origins (or ``photosphere 3D-width'') and primary energy --- are fitted to the pixel contents of 
the different images. For gamma-ray showers, the photosphere 3D-width is found 
to scale with the slant depth of shower 
maximum, an effect related to the variation of the Cherenkov threshold with the
altitude; this property allows to define a dimensionless quantity $\omega$
(the ``reduced 3D-width''), which
turns out to be an efficient and robust variable to discriminate 
gamma-rays from primary hadrons. In addition, the $\omega$ distribution varies only slowly with
the gamma-ray energy and is practically independent of the zenith angle. The
performance of the method as applied to \hess~is presented. Depending on the requirements imposed to
reconstructed showers, the angular resolution at zenith varies from 0.04$^{\circ}$ to 0.1$^{\circ}$
and the spectral resolution in the same conditions from 15\% to 20\%.\newline


PACS-2003: 95.55.Ka, 95.75.-z\\ 

\end{abstract}
\end{frontmatter}

\section{Introduction}
Important progress has recently been achieved in gamma-ray astronomy above 100~GeV 
due to the performance of new stereoscopic systems of Atmospheric Cherenkov Telescopes 
(ACT's)
equipped with high-definition imaging cameras, such as CANGAROO-3
\cite{cangaroo}, \hess
\cite{hofmann} and VERITAS \cite{veritas}. 
These setups bring significant improvements in flux
sensitivity, angular resolution and energy resolution, with respect to previous 
experiments in ground-based gamma-ray astronomy. 
The increase in sensitivity is directly related to
the capability of rejecting hadron-induced showers on the basis of the image
shapes and, at least for known point-like sources, of shower directions. 
Most presently available results from ACT's
were obtained with this last constraint, the signal being extracted from the
distribution of the pointing angle (the so-called ``$\alpha$ or $\theta^2$
plots''). However, stereoscopic systems now provide images of
extended sources --- e.g. supernova remnants \cite{nature} --- and discover unexpected 
sources \cite{hotspot}; 
in such studies, hadronic showers must be rejected 
{\it without using the shower direction}. In the present article, we 
refer to this context as well as to that of point-like sources.

The stereoscopic observation directly provides a simple geometrical
reconstruction method based on the Hillas parameters \cite{hillas} from the
different images; this method has been applied in the HEGRA experiment 
\cite{Daum97} and is
presently adapted to \hess \cite{wystan}. High-definition
imaging provides additional constraints which have already proved very useful,
even with a single telescope, as in the CAT experiment~\cite{LeBohec98}; 
in particular, the
longitudinal light profile in the image can be modeled, providing a likelihood
parameter discriminating gamma-rays from hadrons as well as a 
simultaneous determination of the direction of the primary $\gamma$-ray, 
of the impact parameter and of the shower energy. This method was 
recently extended to \hess, both in
single and multi-telescope modes \cite{de Naurois03}. 

The method described in this article is based on a simple 3D-modeling of 
the Cherenkov light emitted by an electromagnetic air shower. The 
rich information contained in several fine-grained images of such a shower
provides enough constraints to allow an accurate reconstruction, even by means of
a simple model incorporating the rotational symmetry of the
electromagnetic cascade with respect to its incident direction. This allows
to select $\gamma$-ray-induced showers on the basis of only two criteria 
with a direct physical meaning: rotational symmetry and small lateral spread.
In section \ref{sec-recmeth}, the simple assumptions used in the shower 
model are described and justified. In section \ref{sec-discri}, on the basis of 
simulated gamma-rays and of real \hess~data from the blazar PKS~2155-304, 
it is shown that a single variable $\omega$, directly related to the lateral 
spread of Cherenkov photon origins, 
is an ideal parameter for discriminating primary $\gamma$-rays
from hadrons. In section \ref{sec-perfo}, the performance of the method in terms
of gamma-ray reconstruction efficiency, of angular resolution and of 
hadronic rejection is evaluated. Finally, the spectral resolution is discussed.
\section{The reconstruction method}
\label{sec-recmeth}
\subsection{Modeling assumptions}
\label{sec-model}
\begin{figure}
\epsfysize=7cm
\leavevmode
\centering
\epsfbox{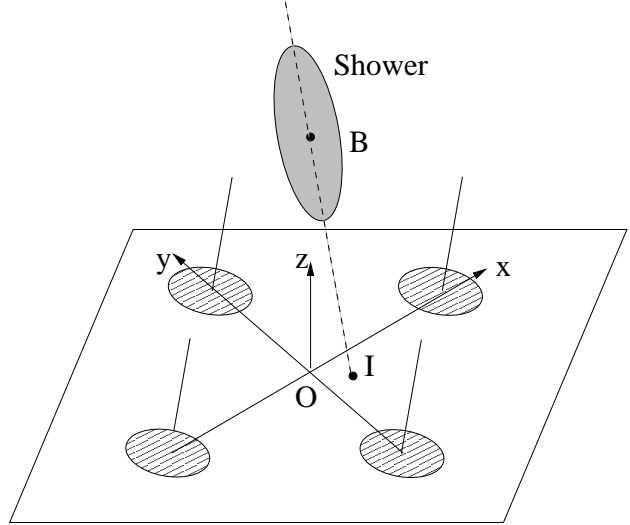}
\caption{\it An electromagnetic shower in the system reference frame.}
\label{fig:system}
\end{figure}

In order to predict the distribution of Cherenkov light in the cameras of the different
telescopes as expected from a $\gamma$-ray shower, one essentially needs 
the spatial distribution of the emission points of Cherenkov photons,
and the angular distribution of these photons with respect to the shower
axis. The most important characteristics of electromagnetic showers is that the 
distributions of secondary particles are on average rotationally symmetric with 
respect to the shower axis. The 3D-model of $\gamma$-ray showers presented here is 
based on the simplifying assumptions indicated below:
\begin{enumerate}
\item The emission points of Cherenkov photons are distributed according to a
3-dimensional Gaussian law with rotational symmetry with respect to the shower
axis, thus characterized by the following parameters 
(figure~\ref{fig:system}):
\begin{itemize}
\item the polar angles $\theta_0$ and $\phi_0$ of the shower axis in the reference 
frame of the stereoscopic system;
\item the coordinates $x_0$ and $y_0$ (in the same frame) of its intersection I 
with the ground (``shower core position on the ground'');
\item the position of the barycentre B (``shower maximum'') on the axis, given
by the distance $h=IB$;
\item the longitudinal ($\sigma_L$) and two-fold degenerate transverse ($\sigma_T$) standard 
deviations of the Gaussian distribution, referred to as ``3D-length'' 
and ``3D-width'' respectively;
\item the total number $N_c$ of Cherenkov photons emitted by the shower.
\end{itemize}
The preceding quantities, referred to as ``shower parameters'', will be
determined by the maximum likelihood fit described in section \ref{sec-fit}.
\item The angular distribution of Cherenkov photons, with respect to the shower
axis, is assumed to be independent of the position of the emission point and of
the energy of the primary $\gamma$-ray; its form, which will be given later, 
is characterized by a single parameter depending on the zenith angle only.
\end{enumerate}
\subsection{Discussion of the preceding assumptions}
\label{sec-ass}
\subsubsection{Assumption (1)}
\begin{figure}
\begin{minipage}[t]{0.47\linewidth}
\epsfig{file=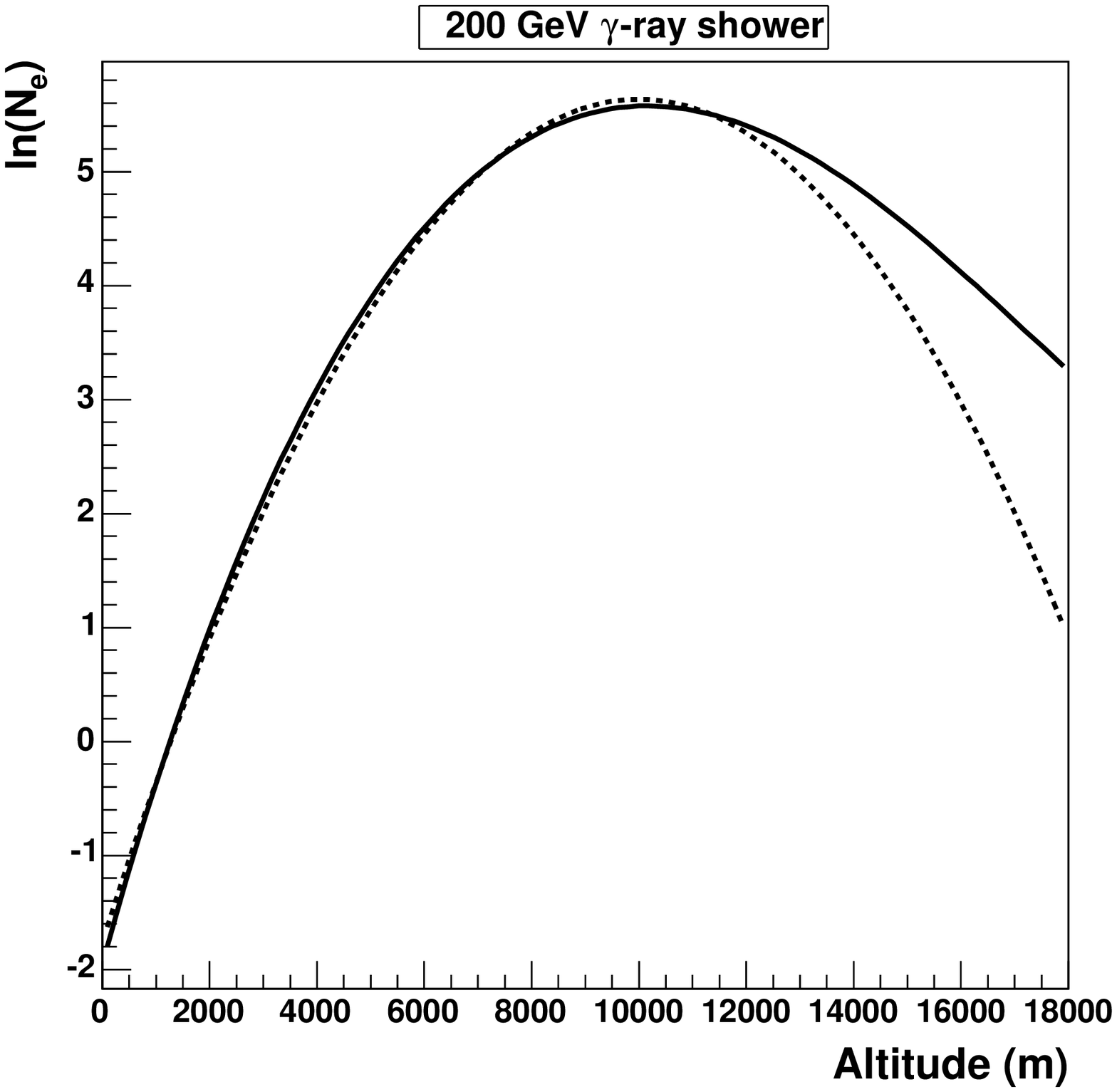,width=\linewidth}
\caption{\it Average longitudinal profile of a 200~GeV vertical $\gamma$-ray shower:
the logarithm of the number $N_e$ of electrons and positrons is plotted
as a function of altitude (solid line). The approximation of the distribution
of $N_e$ by a Gaussian is shown by the dotted line.}
\label{fig:gauss}
\end{minipage}
\hspace{0.05\linewidth}
\begin{minipage}[t]{0.47\linewidth}
\epsfig{file=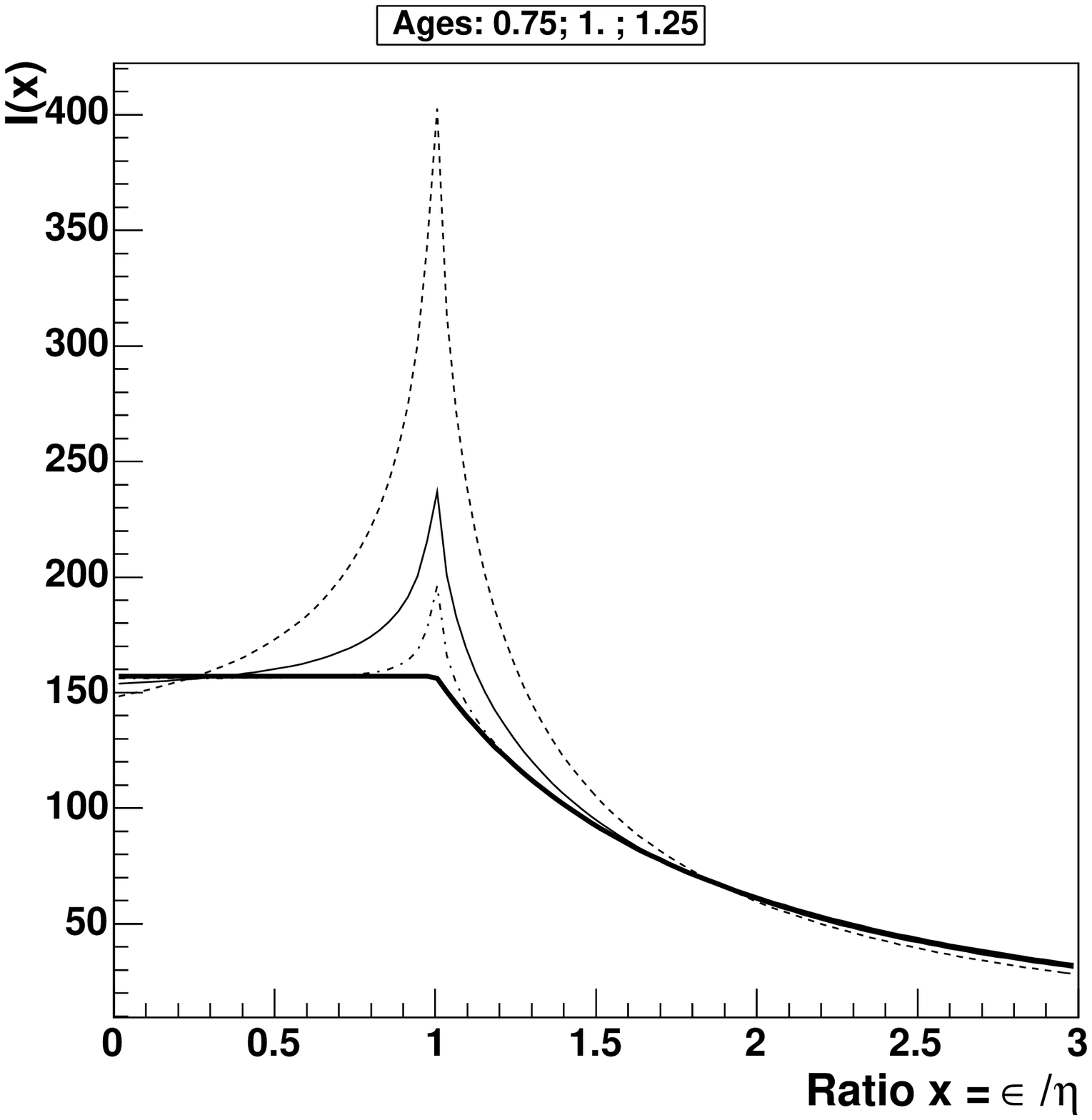,width=\linewidth}
\caption{\it Probability of emission per unit solid angle of Cherenkov 
photons as a function of the angle $\varepsilon$ with respect to the shower axis. 
The distribution is shown for a shower in the 500~GeV-1~TeV energy range at 
three different levels
of its development, corresponding to shower ages of 0.75, 1.0 and 1.25.
The abscissa is the ratio $x=\varepsilon/\eta$, where $\eta$ is the
maximal Cherenkov angle at the altitude considered. The graph of the function 
$I(\varepsilon)$ used in the reconstruction procedure is shown by the bold line.}
\label{fig:tcher}
\end{minipage}
\end{figure}

The average longitudinal development of an electromagnetic shower, i.e. the
number of electrons and positrons as a function of altitude is given by
the Greisen formula \cite{Gaisser92a} and the density profile of the atmosphere.
The number of Cherenkov photons emitted per unit length along the
shower axis follows a similar distribution; the variation of
the Cherenkov threshold with altitude results in a slight shift of about 0.3
radiation lengths downwards of the Cherenkov profile with respect to the
$e^{\pm}$ profile. The average longitudinal profile of a 200~GeV vertical
$\gamma$-ray shower is shown in figure~\ref{fig:gauss}. This profile is well described by a Gaussian
distribution (dotted line in figure~\ref{fig:gauss}) for altitudes lower than
12000~m, particularly in the vicinity of the shower maximum which gives the
dominant contribution to the number of collected Cherenkov photons. As a matter
of fact, the part of the shower above 12000~m only contributes a few percent
of collected photons. Assumption (1) is thus well justified as far as the
longitudinal profile is concerned; the average 3D-length obtained from 
profiles such as the one shown in figure~\ref{fig:gauss} is of the order of
3000~m at zenith, almost independent of the primary energy.

On the other hand, the photosphere 3D-width $\sigma_T$ represents the average lateral spread of 
the origins of Cherenkov photons. The lateral profile is assumed to be Gaussian;
this is not exact since
the real distribution is more sharply peaked close to the axis; however,
it will be shown (from simulations as well as from
genuine~\hess~data) that the typical 3D-width of electromagnetic showers 
at zenith is of the order of 10~m, and in
most cases smaller than 15~m. Such a structure at a typical distance of 10~km
from the telescope is viewed under an angle of 1.5~mrad, smaller than the pixel
size. Thus, there is no need for a more accurate description of the lateral
distribution. The photosphere 3D-width, obtained with the constraint of rotational
symmetry, is particularly useful
to distinguish $\gamma$-ray from hadron showers which, in most cases, are much 
broader.
\subsubsection{Assumption (2)}
A priori, assumption (2) looks more difficult to justify, since the angular 
distribution of Cherenkov photons with respect to the shower axis results from
that of electrons (which depends upon the position within the shower) combined
with the Cherenkov cone of each electron whose half-angle depends on the local
atmospheric density. It is thus appropriate to use the variable 
$x=\varepsilon/\eta$, in which $\epsilon$ is the angle between the direction of 
the Cherenkov photon and the shower axis, and 
$\eta$ is the maximal Cherenkov angle at the altitude of
the emission point ($\cos \eta =
1/n(z)$, $n(z)$ being the refraction index at altitude~$z$).
The lateral position in the shower should not be essential
for the reasons explained above; on the other hand, the longitudinal position
can be characterized by the shower age, as defined in 
reference~\cite{Gaisser92a}. For a given age, one is interested in the probability 
$I(\varepsilon)$ of emission of a Cherenkov photon per unit solid angle around
a direction making an angle $\varepsilon$ with the shower axis. The normalization is
defined by:
\[ \int I(\varepsilon) \, d\Omega \approx 2 \pi \int I(\varepsilon) \: 
\varepsilon  \, d\varepsilon = 1 \: . \]
Simulations show that, expressed as a function of $x = \varepsilon/\eta$,
the functions $I$ at fixed age do not depend  strongly on
the initial energy. Some of them are shown in figure \ref{fig:tcher} for 
showers in the 500~GeV-1~TeV energy range at different ages. The distributions 
are similar except for the peak at $x=1$, which rapidly decreases when 
the shower develops. Taking this peak into account in the 3D-model did not 
improve the results significantly. Therefore, we use the following universal function $I(x)$:
\[ I(\varepsilon) = K \: \: \: \mbox{if} \: \: \: \varepsilon < \eta \]
\[ I(\varepsilon) = K \, \frac{\eta}{\varepsilon }
\exp \left[ - \frac{\varepsilon-\eta}{4 \eta} \right] 
\: \: \: \mbox{if}\: \: \:  \varepsilon > \eta \]
in which $K$ is fixed by the normalization:
$K = 1/( 9 \pi \eta^2 )$. The graph of this function (the thick line 
in figure \ref{fig:tcher}) is in reasonable agreement with the curves obtained from 
simulations, except
in the vicinity of $\varepsilon = \eta$. This simplified form actually
neglects the contribution of the narrow peak around 
$\varepsilon = \eta$ which is only important for small ages, i.e. for
high altitudes. Ideally, $\eta$ should
be calculated at the altitude of shower maximum, which is a parameter of the likelihood fit.
However, coupling the angular distribution of Cherenkov light to the altitude often prevents the
maximization procedure from converging and it turned out to be sufficient to use a 
value of $\eta$ independent of the altitude and of the energy. Its dependence 
on the zenith angle~$\zeta$ was chosen as $\eta= 15~\mbox{mrad} \sqrt{\cos \zeta}$, 
a relation which follows if the atmospheric density is assumed to decrease exponentially with
altitude.
\subsection{Implementation of the method}
\label{sec-fit}
\begin{figure}
\epsfysize=7cm
\leavevmode
\centering
\epsfbox{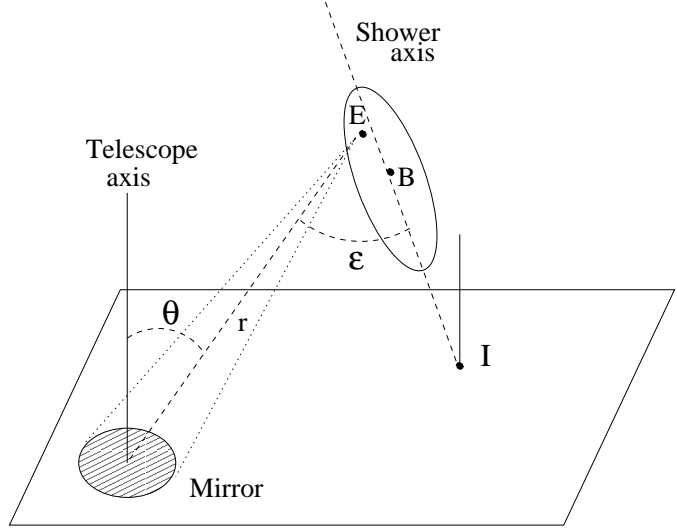}
\caption{\it Calculation of the expected number of collected photons in a pixel, as a
function of the shower parameters.}
\label{fig:tele}
\end{figure}

The preceding 3D-model enables us to work out the expected number of Cherenkov photons
$q_{th}$ collected by a given pixel of a given telescope as a function of the shower 
parameters listed in section~\ref{sec-model}. The calculation is done in the
reference frame of the telescope of interest as shown in figure \ref{fig:tele}. 
The average direction of observation of the pixel of interest makes an angle 
$\theta$ with the telescope axis and an angle $\varepsilon$ with the shower axis.
We have to sum up the contributions of all photons emitted within the solid angle
$\Delta \omega_{pix}$ covered by the pixel, provided they are pointing towards the 
telescope mirror of area $S_{tel}$. At the distance $r$ from 
the telescope along the line of sight, in a volume $r^2 dr \Delta \omega_{pix}$, the density
of Cherenkov photons $n_c(r)$ is obtained from the shower parameters, using 
assumption~(1). From an emission point $E$ in this volume, the mirror is viewed 
under the solid angle $d \Omega = S_{tel} \cos \theta/ r^2$ and the fraction of 
those photons reaching the mirror is given by $I(\varepsilon) S_{tel} 
\cos \theta / r^2$.  Integrating along the line of sight, one finds the expected
value of the number of photons collected by the pixel of interest:
\begin{eqnarray}
 q_{th} & = & \int_0^{\infty} n_c(r) \, r^2 dr \, \Delta \omega_{pix} \: 
I(\varepsilon) \, \frac{ S_{tel} \cos \theta}{r^2 } \nonumber \\  
& = & S_{tel} \, \Delta \omega_{pix} \, 
I(\varepsilon) \, \cos \theta  \,  \int_0^{\infty} n_c(r) \, dr  
\label{eq:qth}
\end{eqnarray}
The last integral is easily calculated, due to assumption~(1), as shown in Appendix 1.

The quantities $q_{th}$ for all pixels, calculated for a set of shower parameters,
are further used to build up a likelihood function for each event including at least two
images of a given shower. Images have been previously submitted to a cleaning procedure
in order to remove isolated clusters of pixels with a small charge which likely result 
from the night sky background. In the case of \hess~data analyzed in this article, the
cleaning criteria were similar to those of reference \cite{wystan} but slightly
looser: pixels in the image were required to have a charge content above  
5~photoelectrons and to have a neighbour above 7~photoelectrons; conversely, 
pixels above 7~photoelectrons were required to have a neighbour above 5~photoelectrons;
otherwise, their charge contents were cleared.
Pixels retained by this cleaning procedure, as well as their 
immediate cleared neighbours, were further used in the likelihood calculation; on the other hand, 
pixels invalidated by the calibration procedure were not included. 
The fluctuations considered in the fit
are Poisson fluctuations on the predicted number of photons falling onto a pixel
and those due to the phototube responses, the latter being considered as Gaussian. 
Fluctuations in the shower development are not taken into
account, nor are correlations between the contents of different pixels. 
This has little effect on the following analysis since no goodness-of-fit parameter is used
to discriminate gamma-ray showers from hadronic ones. 
The likelihood function for each event, described in Appendix 2,
is then maximized with respect to the 8 shower parameters defined in section~\ref{sec-model}. 
This is achieved by means of the MINUIT program from CERN \cite{minuit}. 
The simple assumptions of the 3D-model allow a rather fast processing of the 
events.

It should be emphasized that the present 3D-reconstruction of showers is relatively insensitive 
to the presence of invalidated pixels in cameras; for example, an extreme situation in which, 
in all cameras, 20$\%$ of the pixels would be damaged would mainly result in a degrading of the 
angular resolution (an increase from 0.12$^\circ$ to 0.15$^\circ$ of the angular error) with only 
a 10$\%$ loss in reconstruction efficiency.

\section{Gamma-ray/hadron discrimination based on shower shape}
\label{sec-discri}
\subsection{Longitudinal development: physical condition for gamma-rays}
\begin{figure}
\epsfig{file=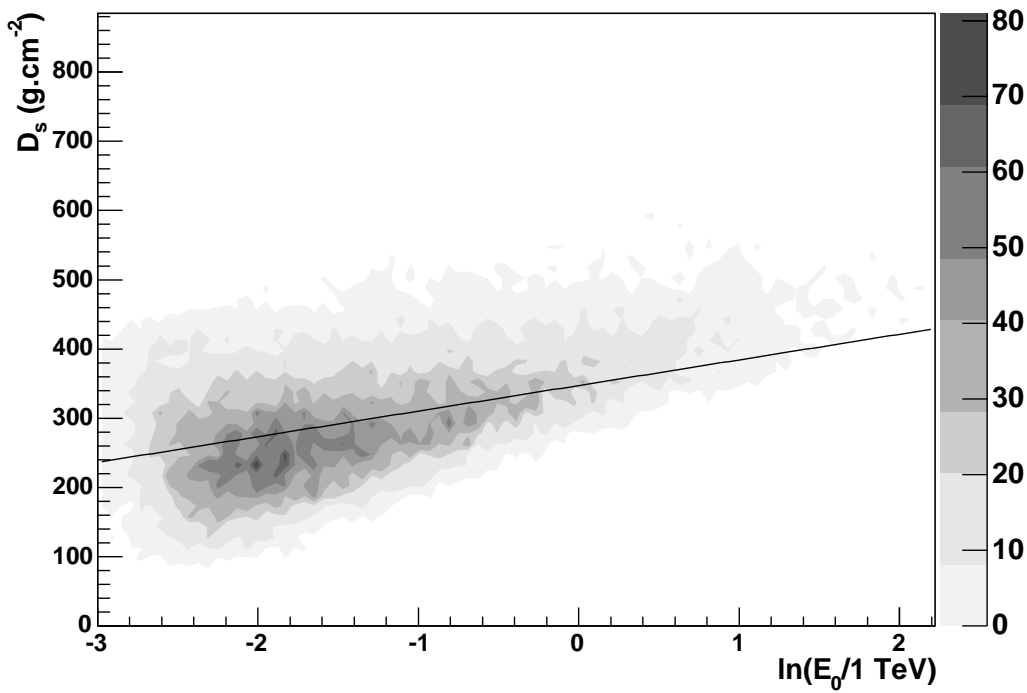,width=\linewidth}
\caption{\it Gamma-ray showers simulated at zenith with a differential energy spectrum proportional
to $E_0^{-2.2}$: depth of shower maximum $D_s$ reconstructed by the fit, as a function of $\ln E_0$.
The line shows the relation between the average value of $D_s$ and $\ln E_0$ given by the Greisen formula~\cite{Gaisser92a}.}
\label{fig:altgreisen}
\end{figure}
\begin{figure}
\begin{minipage}[t]{0.47\linewidth}
\epsfig{file=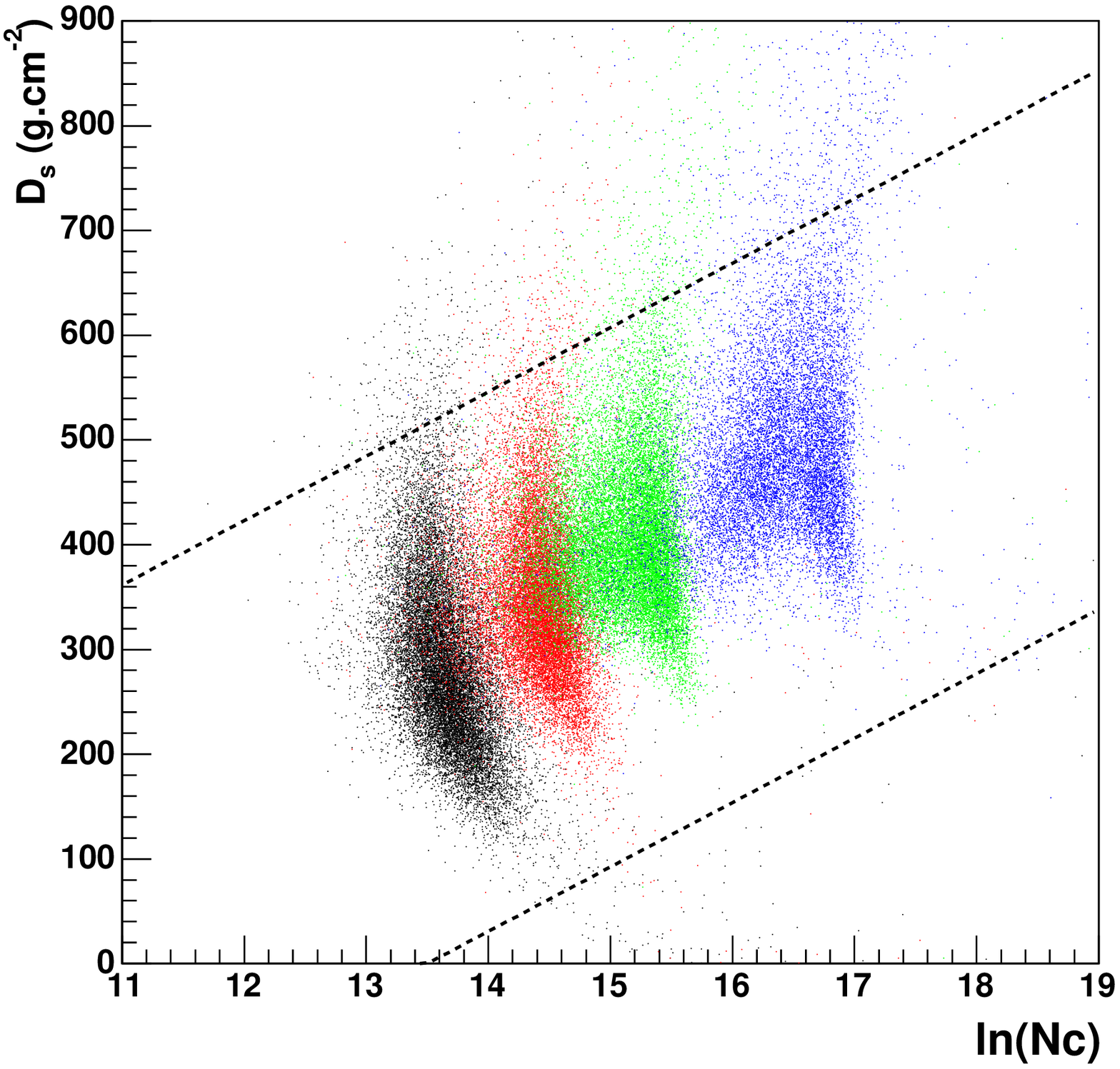,width=\linewidth}
\caption{\it Simulated gamma-ray showers: depth of shower maximum $D_s$ 
as a function of the logarithm of the total number $N_c$ of Cherenkov photons
(both reconstructed by the fit), for different values of
the primary energy $E_0$ (from left to right: 200~GeV, 500~GeV, 
1~TeV, 5~TeV)
at zenith. Condition (\ref{eq:physreg}) corresponds to the region between the 
two straight lines; the upper bound is more restrictive for $\gamma$-rays, due to the
accumulation of background events at large depths.}
\label{fig:nc-vs-depth}
\end{minipage}
\hspace{0.06\linewidth}
\begin{minipage}[t]{0.47\linewidth}
\epsfig{file=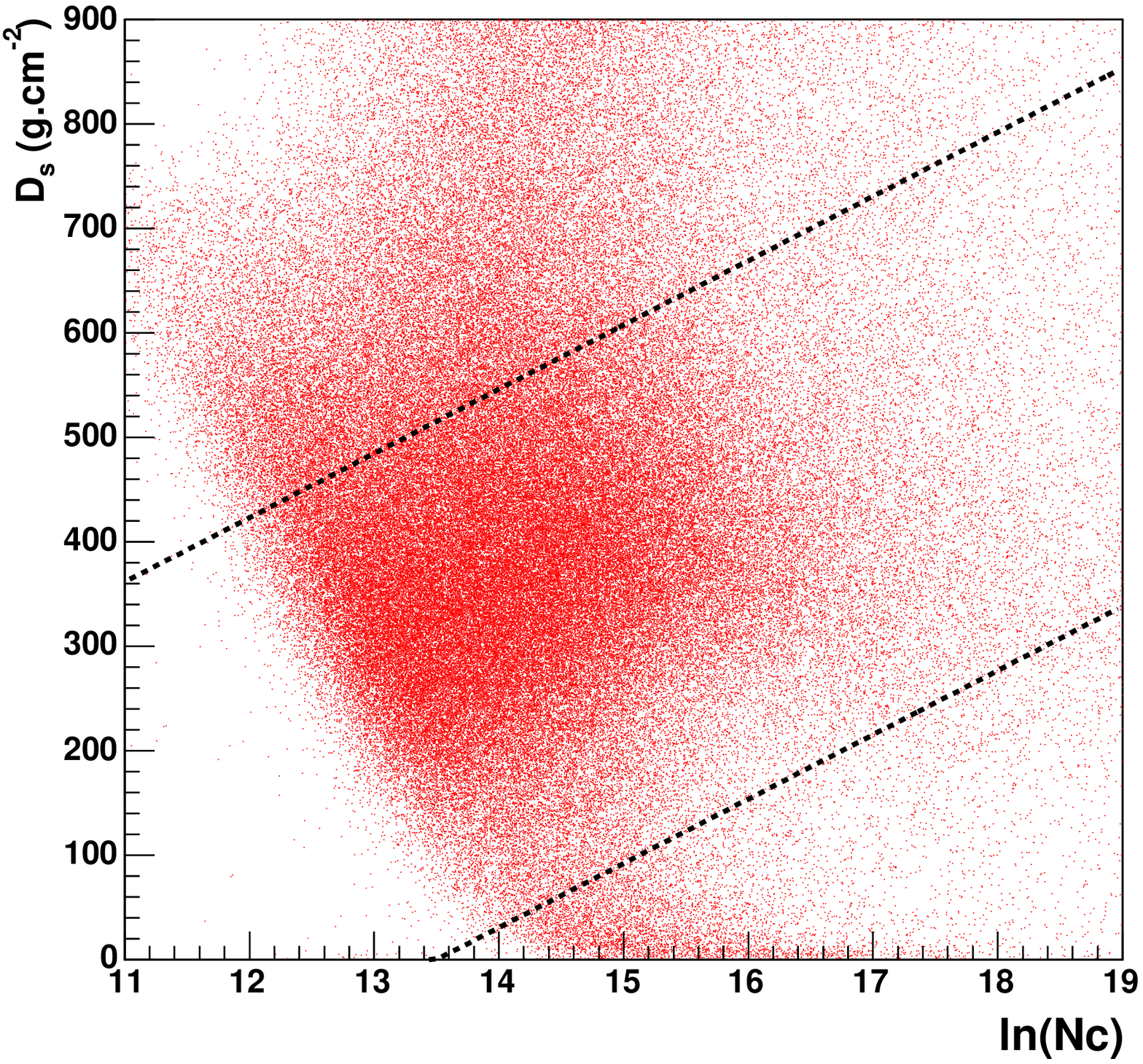,width=\linewidth}
\caption{\it Showers from a field of view free of $\gamma$-ray source (far off the Galactic
Plane) reconstructed as if they were electromagnetic showers: depth of shower 
maximum $D_s$ as a function of the logarithm of the total number $N_c$ of Cherenkov photons
(both reconstructed by the fit). The observation was performed at a
zenith angle of 16$^{\circ}$.
Condition (\ref{eq:physreg}) corresponds to the region between the 
two straight lines.}
\label{fig:nc-had}
\end{minipage}
\end{figure}
Simulations of gamma-ray-induced showers at different energies and zenith angles
were used to produce images in the \hess~telescopes under normal experimental
conditions (including light pollution by the night sky background). Those images were
submitted to the cleaning procedure and analyzed according to the 3D-model. 
About 90\% of the simulated showers were successfully reconstructed.
Since only one length scale, namely the radiation length, 
governs the development of electromagnetic showers both longitudinally and 
laterally, it is convenient to express characteristic lengths in units of 
radiation lengths (or equivalently in ``air thickness'' in g~cm$^{-2}$).
The longitudinal development is thus characterized by the slant depth 
$D_s$ of shower maximum (i.e. the thickness of air between the top of the 
atmosphere and the maximum, as measured along the shower axis); it is
calculated from the altitude $z_{max}$ of the barycentre (shower maximum) 
obtained by the likelihood fit and from the zenith angle $\zeta$. 
The relevance of the altitude reconstruction is illustrated in figure
\ref{fig:altgreisen} in which $D_s$ is plotted as a function of the 
logarithm of the true energy $E_0$ for a sample of gamma-rays generated with a
differential spectrum proportional to $E_0^{-2.2}$. As expected from the Greisen
formula \cite{Gaisser92a}, the average depth of shower maximum increases linearly 
with $\ln E_0$. A similar correlation is obtained between $D_s$ and 
the logarithm of the total number of Cherenkov photons $N_c$,
both quantities being reconstructed by the fit; this is shown in figure
\ref{fig:nc-vs-depth} for several values of the primary gamma-ray energy $E_0$. 
This property allows to define an additional constraint characterizing gamma-ray showers:
the fitted parameters $D_s$ (in g~cm$^{-2}$) and $N_c$ are required to satisfy
the following condition~:
\begin{equation}
  61.5 \, (\ln N_c -13.5 + T(\zeta)) \leq D_s \: \: \leq  61.5 
  \, (\ln N_c - 10 + T(\zeta)) + 300
\label{eq:physreg}
\end{equation}
with $T(\zeta) = 3.28 (1 - \cos \zeta)$, this last term being introduced since
the relation between the reconstructed value of $N_c$ and the estimated value of
the primary energy depends on $\zeta$. Events satisfying condition 
(\ref{eq:physreg}) are represented in
the region between the two straight lines shown in figure \ref{fig:nc-vs-depth} 
for showers at zenith. The fraction of gamma-rays reconstructed by the fit 
(about 90\%) is almost unaffected when requiring condition (\ref{eq:physreg}) 
which results in an additional loss of 2\% for zenith angles lower than 50\dg.
On the other hand, the fit, as applied to
hadronic showers, sometimes converges towards very small or, more often,
towards very large depths\footnote{This is why 
condition (\ref{eq:physreg}) is more restrictive for $\gamma$-ray
showers at large depths.} (figure \ref{fig:nc-had}). 
By requiring convergence with fitted parameters satisfying 
condition (\ref{eq:physreg}), one removes 70\% of hadronic showers. The 
remaining ones are compatible with rotational symmetry and condition 
(\ref{eq:physreg}); however, their 3D-width distribution, much broader than 
that of gamma-rays, offers additional rejection criteria which we investigate
now. \\
\subsection{Lateral development of the shower}
\subsubsection{The 3D-width as a gamma-ray/hadron discriminating
variable}
\label{sec-3Dw}
\begin{figure}
\begin{minipage}[t]{0.48\linewidth}
\epsfig{file=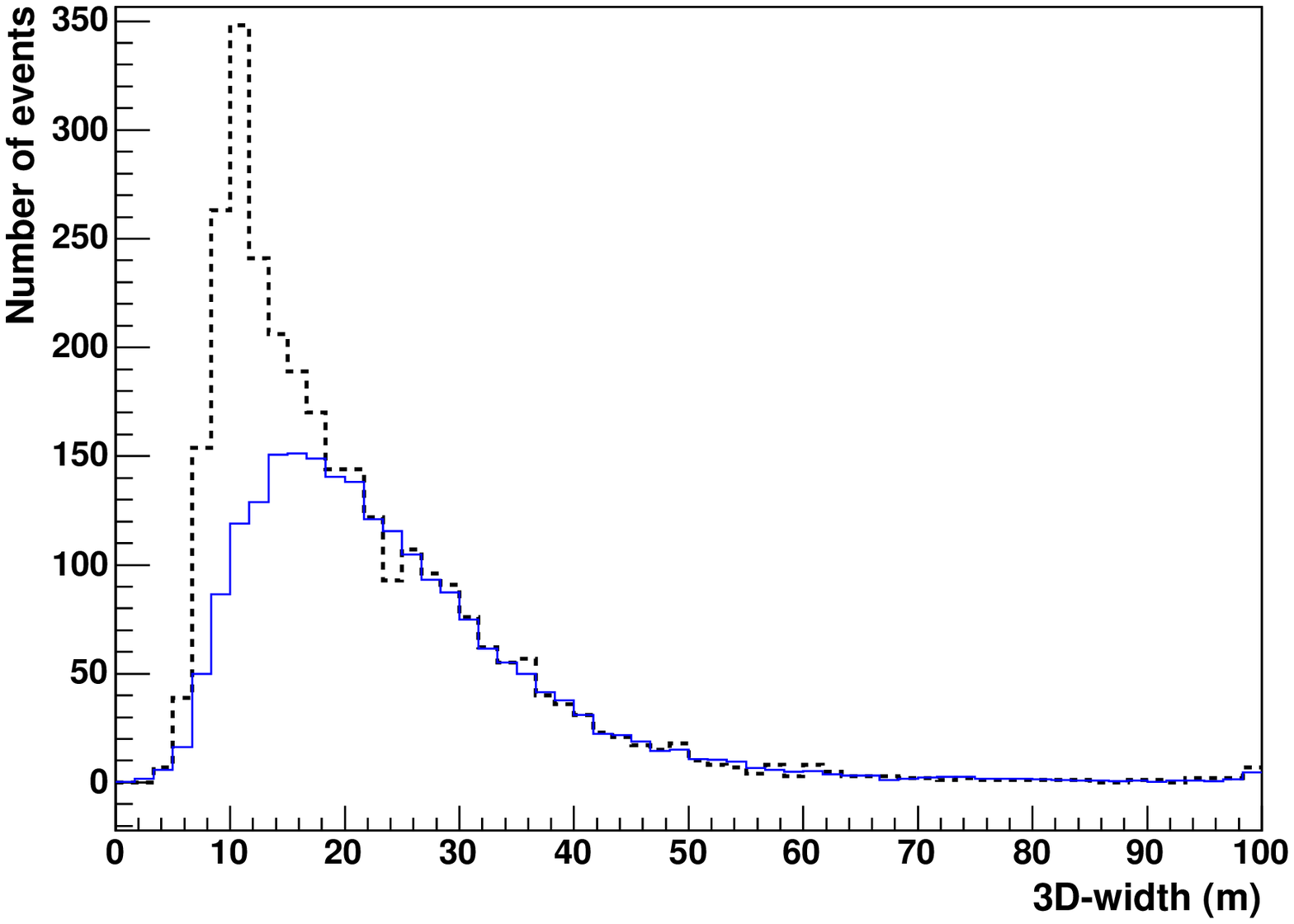,width=\linewidth}
\caption{\it Distribution of the 3D-width from data taken on PKS2155-304
in 2004. (a) Dotted line: source region. (b) Solid line: background from 5 control
regions in the same data sample, rescaled by a factor 1/5.}
\label{fig:pksall}
\end{minipage}
\hspace{0.06\linewidth}
\begin{minipage}[t]{0.48\linewidth}
\epsfig{file=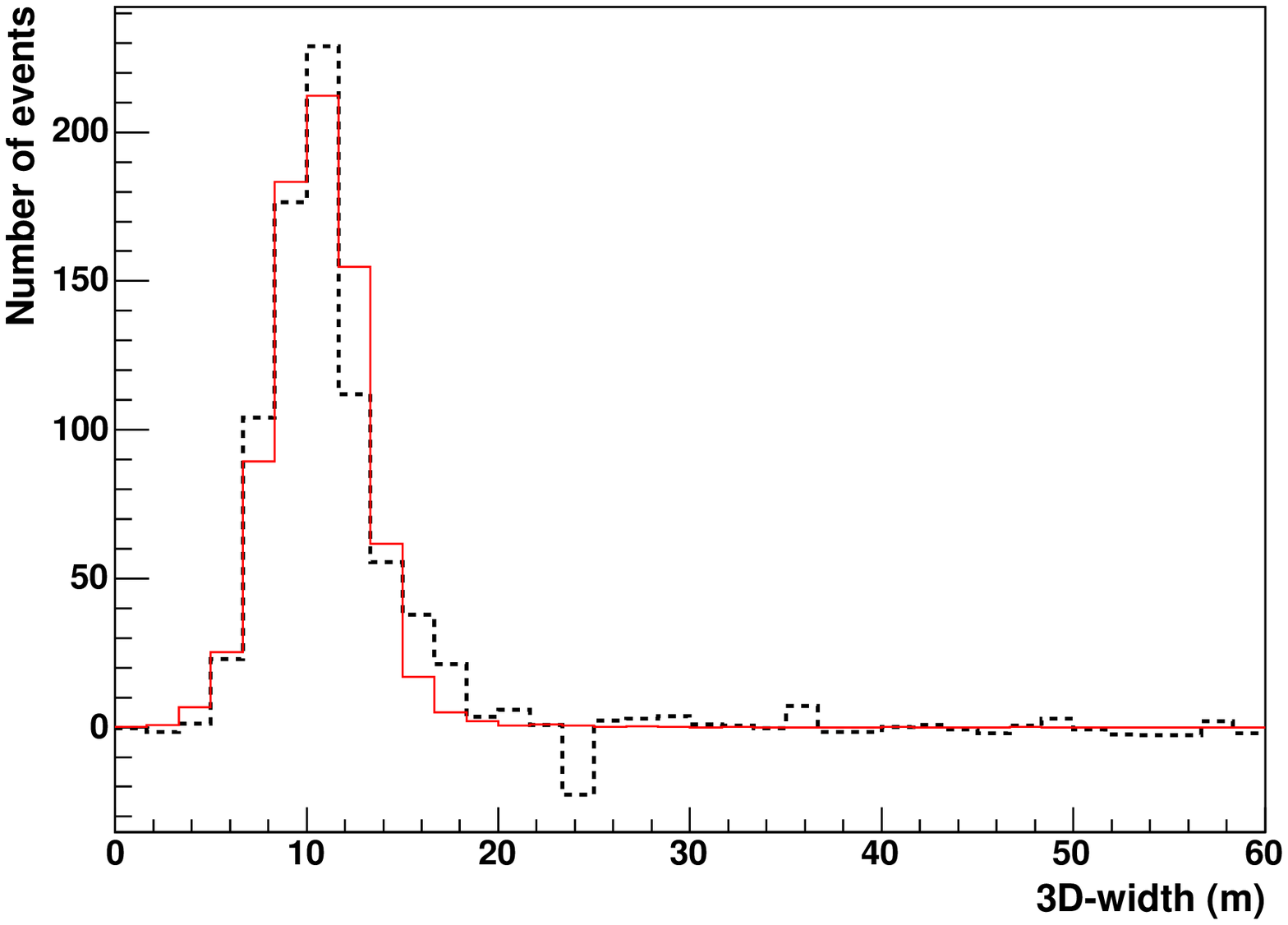,width=\linewidth}
\caption{\it Experimental distribution of $\sigma_T$ for gamma-rays 
(dotted line) compared to that obtained from simulations (solid line).}
\label{fig:pkssim}
\end{minipage}
\end{figure}

\begin{figure}
\begin{minipage}[t]{0.48\linewidth}
\epsfig{file=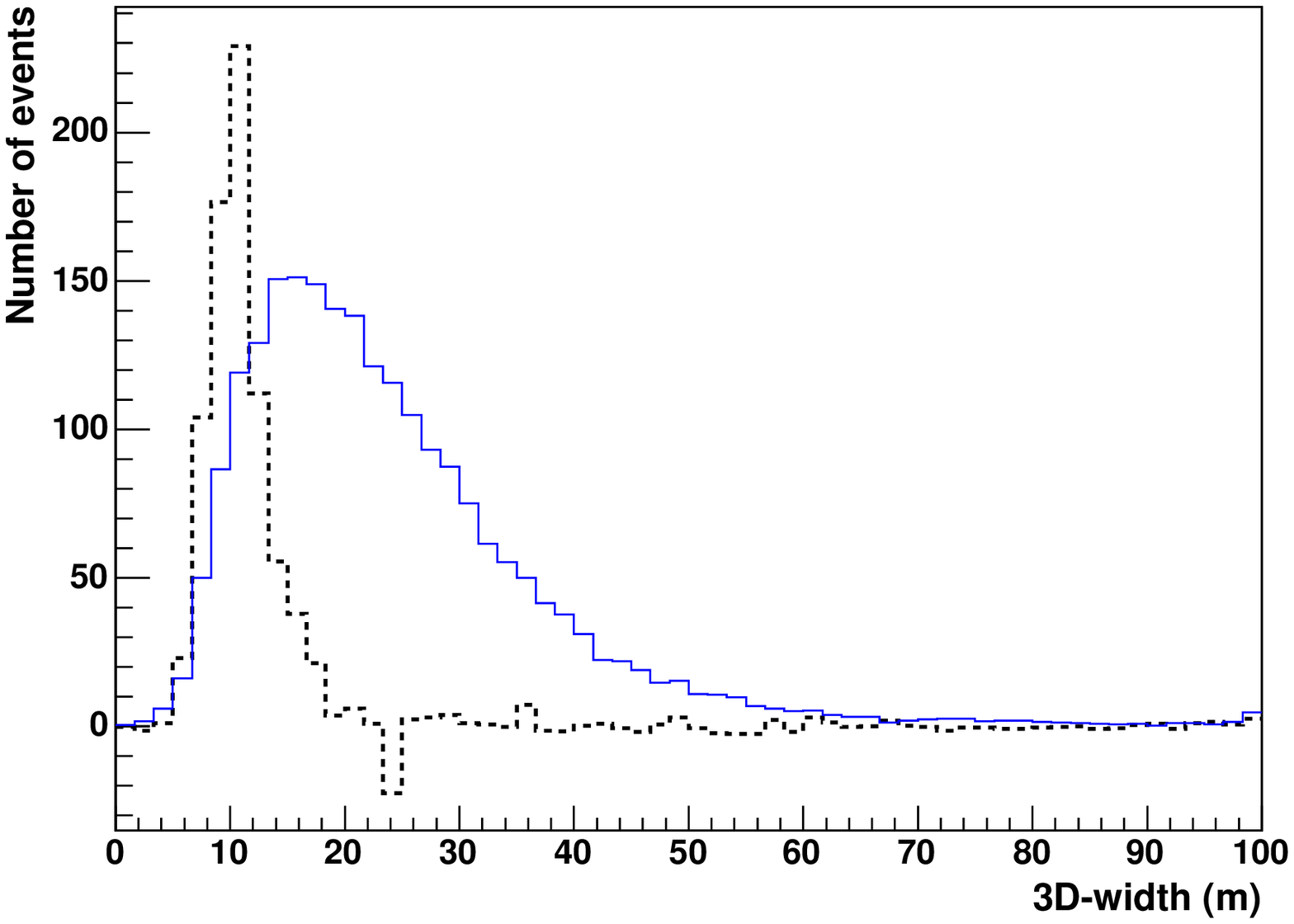,width=\linewidth}
\caption{\it Distribution of $\sigma_T$ for gamma-rays (dotted line) obtained from histograms of figure \ref{fig:pksall} 
by subtracting the background from on-source data; the background distribution
is given by the solid line.}
\label{fig:pkssep}
\end{minipage}
\hspace{0.06\linewidth}
\begin{minipage}[t]{0.48\linewidth}
\epsfig{file=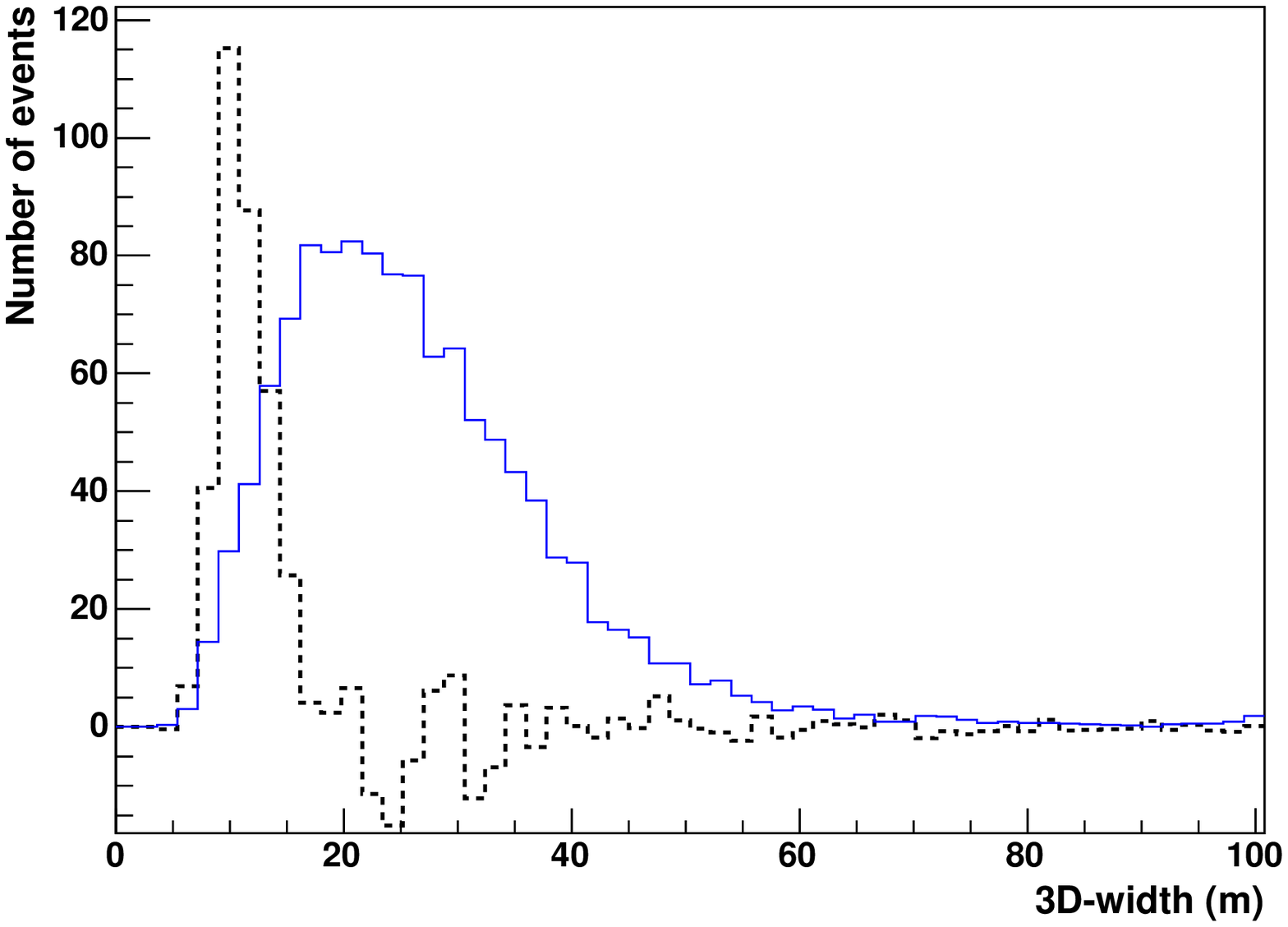,width=\linewidth}
\caption{\it Same histograms as in figure \ref{fig:pkssep} restricted to events
with at least three available images.}
\label{fig:pks3tel}
\end{minipage}
\end{figure}

The potentiality of the 3D-width $\sigma_T$ for discriminating gamma-rays
from hadrons is illustrated in figures \ref{fig:pksall}, \ref{fig:pkssim}, \ref{fig:pkssep},
and \ref{fig:pks3tel}, obtained from the analysis of \hess~data 
(4 hours live time) taken in 2004 on the blazar PKS2155-304 with a mean
zenith angle of 30$^{\circ}$. In order to monitor the hadronic background, data were taken with 
the telescope axis shifted by an angle $\alpha=0.5^{\circ}$ from the position of the source 
(``wobble mode''); in the following, $\alpha$ will be referred to as the offset angle. The
background contribution was obtained from 5 ``off'' regions with the same 
acceptance as the source region. 
Figure \ref{fig:pksall} compares $\sigma_T$ distributions in the source region
(dotted line) and in 5 background control regions (solid line) rescaled by a factor 1/5.
The experimental $\sigma_T$ 
distribution for gamma-rays, obtained by subtraction, is shown in figure
\ref{fig:pkssim} and found to be in very good agreement with that obtained 
from simulations based on the 
spectrum of PKS2155-304 as measured in reference \cite{wystan}. The 3D-widths of
gamma-ray showers are essentially smaller than 20~m, thus contrasting with
those of hadron showers, as shown in figure \ref{fig:pkssep}. The separation
between both populations is further increased by restricting the histogram to
those events observed by at least 3 telescopes (figure \ref{fig:pks3tel}); 
in this case, the corresponding distribution for gamma-rays is almost 
unaffected, whereas that of hadrons is shifted towards higher values. This
is easily understood since $\sigma_T$ has a real physical meaning for gamma-ray
showers and the number of available images only affects the reconstruction error; on the
other hand, for hadrons satisfying the preceding fit, the assumption of
rotational symmetry is generally not valid and the reconstructed value 
depends on the observation conditions. In any case, most hadronic showers 
are found to be much broader than gamma-ray showers.
\subsubsection{The reduced 3D-width}
\label{sec-omega}
Since the preceding fit also yields the altitude of shower maximum, 
it is convenient to 
express $\sigma_T$ in ``air thickness'' in g~cm$^{-2}$, i.e.
to use the quantity $\sigma_T^{\prime} = \sigma_T \: \rho(z_{max})$, in
which $\rho(z_{max})$ is the density of air at the altitude $z_{max}$
of the barycentre. 
\begin{figure}
\begin{center}
\epsfig{file=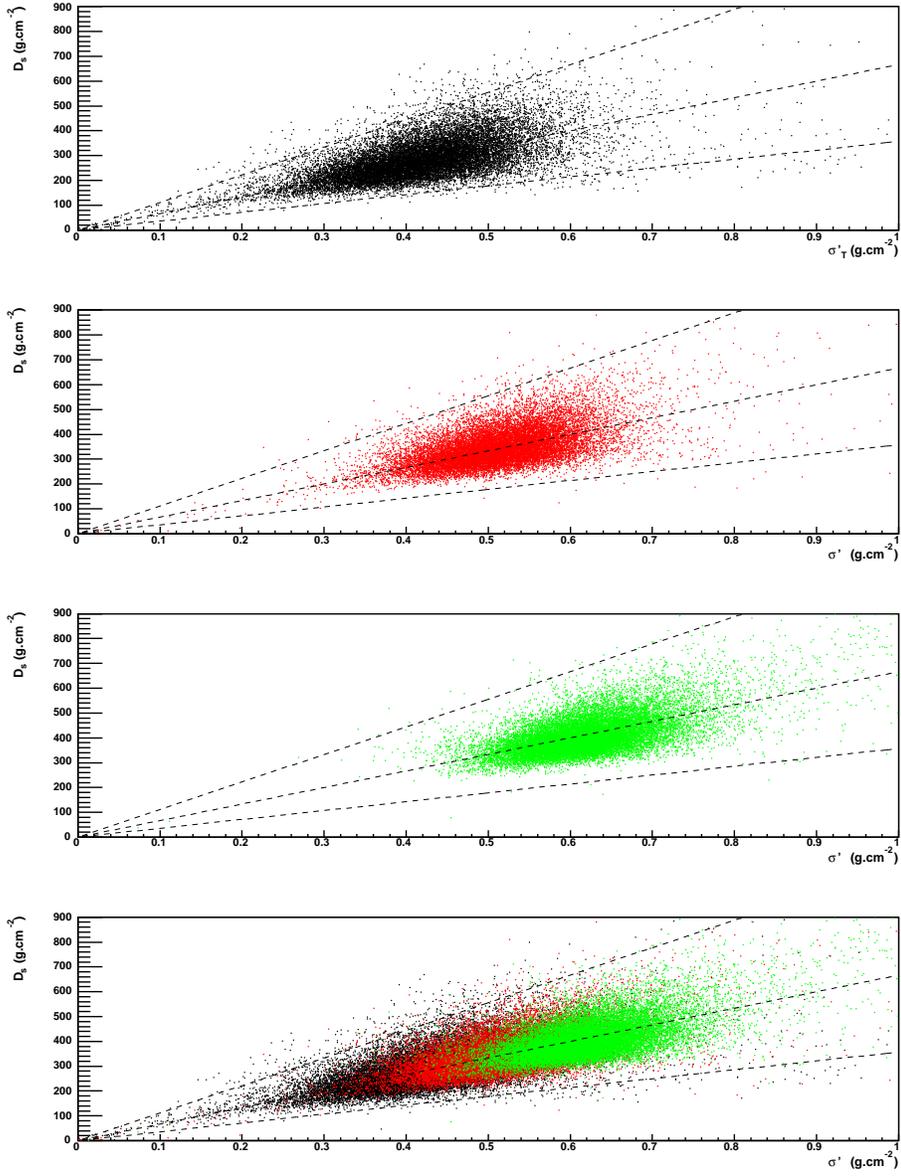,width=\linewidth}
\caption{\it Simulated gamma-ray showers: depth
of shower maximum $D_s$ as a function of the 3D-width $\sigma_T^{\prime}$ 
(both reconstructed by the fit and expressed in g~cm$^{-2}$), for different values of
the primary energy $E_0$: from top to bottom, 200~GeV, 500~GeV, 1~TeV, the last
plot is the superposition of the preceding ones. The straight lines are 
intended to guide the eye.}
\label{fig:w-vs-depth}
\end{center}
\end{figure}
For simulated gamma-ray showers of different primary energies, 
the depth of shower maximum $D_s$
is plotted versus $\sigma_T^{\prime}$ in figure \ref{fig:w-vs-depth}, 
both quantities being reconstructed by the fit and expressed in g~cm$^{-2}$.
The average value of $\sigma_T^{\prime}$ is found to increase with $D_s$. 
This is clearly an effect of the decrease of the Cherenkov threshold with depth; 
at larger depths, more low-energy electrons further from the axis produce 
Cherenkov light. Simulations show that, on average, $\sigma_T^{\prime}$ scales 
with $D_s$ (figure \ref{fig:w-vs-depth}). Consequently, the dimensionless ratio
$\omega = \sigma_T^{\prime}/D_s$ follows a distribution which, in the absence of
reconstruction errors, would be independent of the gamma-ray energy and zenith
angle. In the following, the quantity $\omega$ will be referred to as the ``reduced 
3D-width''. 
\begin{figure}
\begin{minipage}[t]{0.47\linewidth}
\epsfig{file=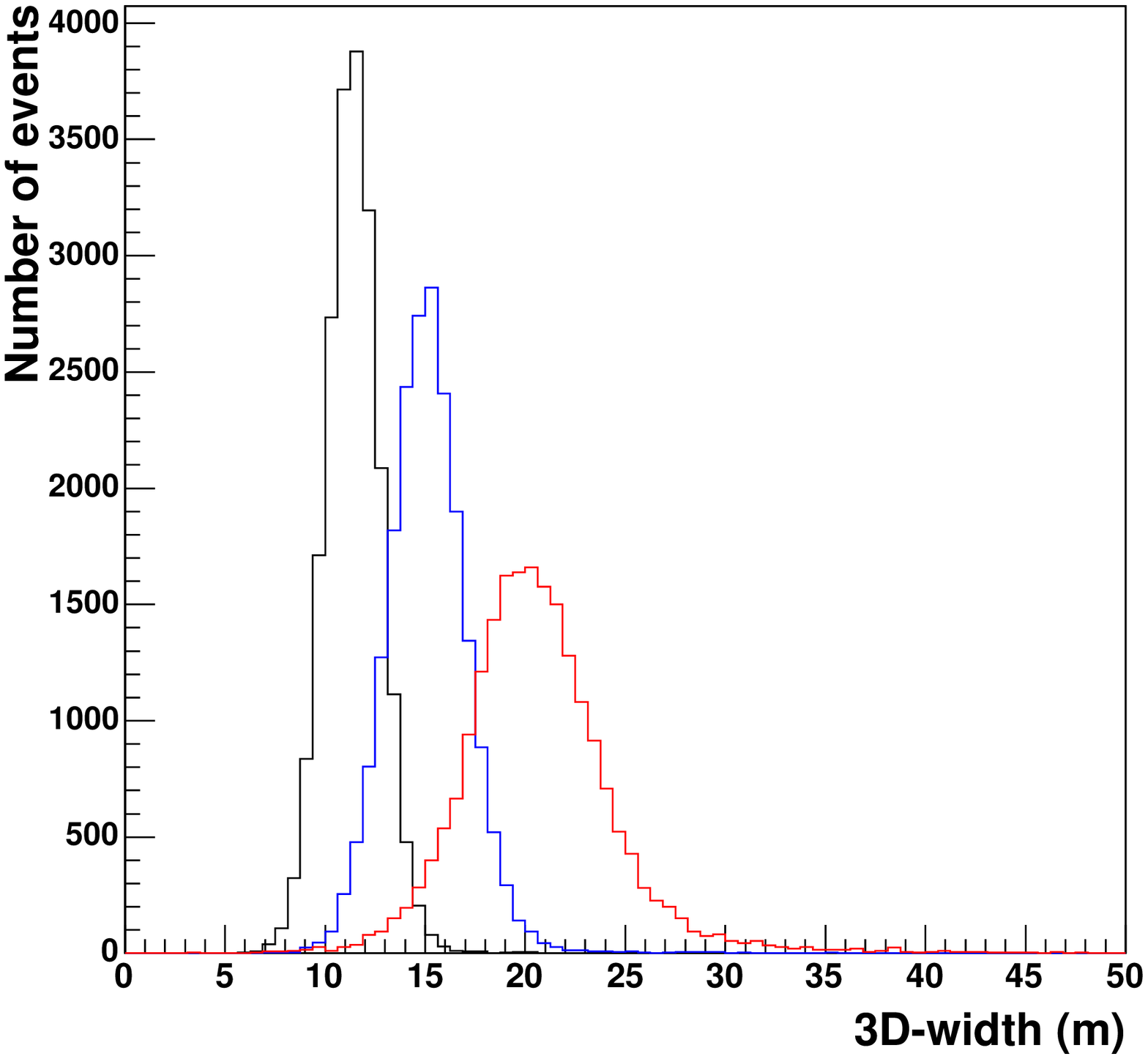,width=\linewidth}
\caption{\it 3D width distributions for 1~TeV gamma-ray showers at zenith angles 
$0^{\circ}$ , $46^{\circ}$ and $60^{\circ}$ (from left to right).}
\label{fig:zenwidth}
\end{minipage}
\hspace{0.06\linewidth}
\begin{minipage}[t]{0.47\linewidth}
\epsfig{file=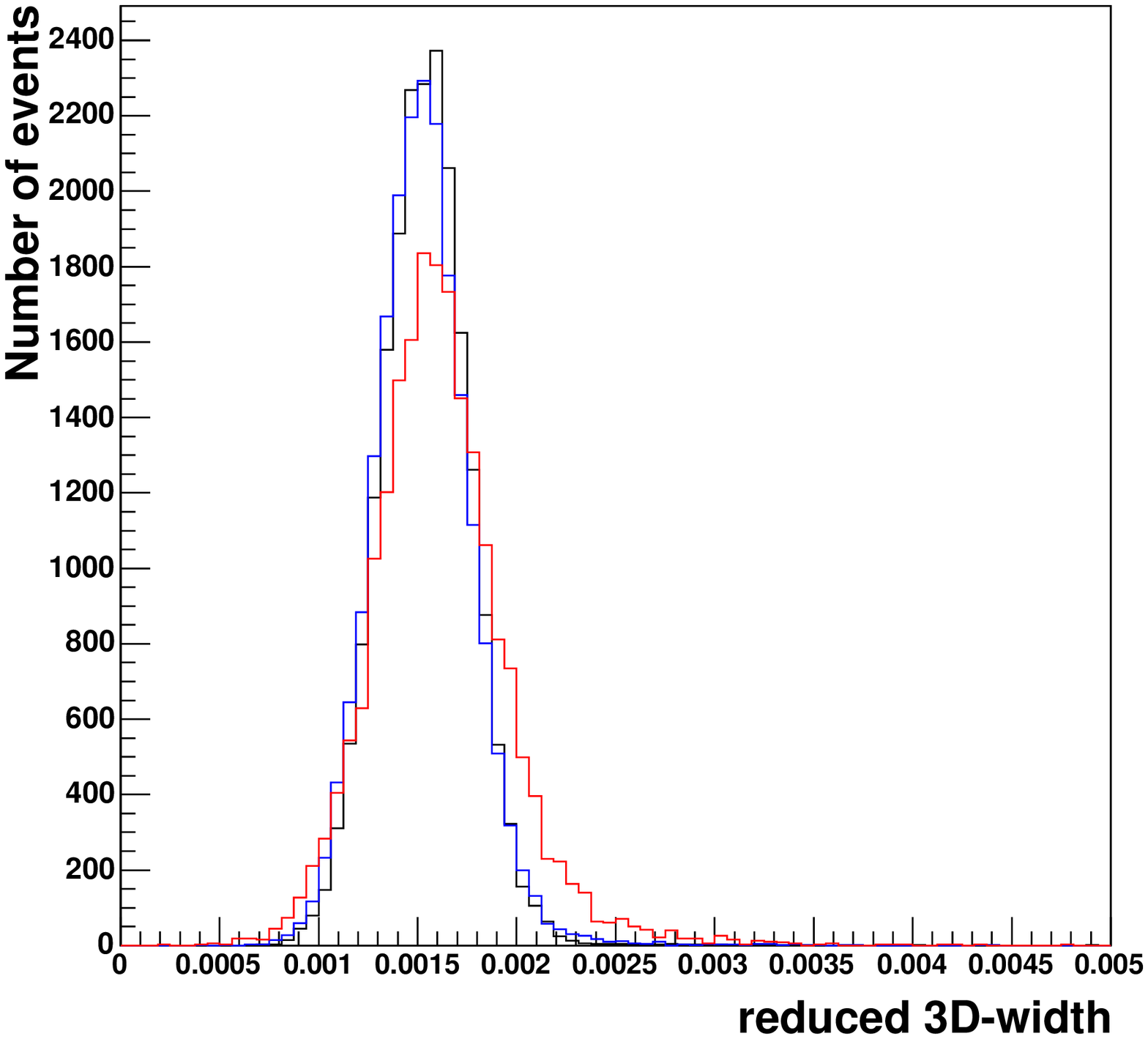,width=\linewidth}
\caption{\it Distributions of the reduced 3D-width 
at zenith angles $0^{\circ}$ , $46^{\circ}$ and $60^{\circ}$ for 1~TeV
$\gamma$-ray showers.}
\label{fig:zenred}
\end{minipage}
\end{figure}
Figure \ref{fig:zenwidth} shows the $\sigma_T$ distributions for 1~TeV gamma-ray
showers at zenith angles $0^{\circ}$ , $46^{\circ}$ and $60^{\circ}$; in
contrast, figure \ref{fig:zenred} shows the corresponding distributions for the
reduced 3D-width $\omega$, almost identical as can be seen from table
\ref{tab:zen}; only a small extension towards higher values appears at large zenith
angles. Furthermore, these distributions depend only slightly on the gamma-ray
energy, as shown in table \ref{tab:en}, the average value increasing linearly
with energy. Consequently, for a large range of primary energies,
the $\omega$ distribution for gamma-rays is essentially concentrated in the
region 
\begin{equation}
0.8 \times 10^{-3} < \omega < 2 \times 10^{-3}  
\label{eq:widthint}
\end{equation}
Of course, the upper
limit is the most relevant for eliminating hadrons, the lower limit being
nevertheless useful for events viewed by 2 telescopes only (see figure
\ref{fig:compgh}).
\begin{table}
\begin{minipage}[t]{0.47\linewidth}
\caption{\it Average values and standard deviations of the $\omega$ distributions 
for 1~TeV gamma-rays.}
\vspace{0.2cm}
\begin{center}
\begin{tabular}{ccc} \hline
Zenith  &  $< \omega >$     & $\sigma(\omega)$ \\ 
 angle  &  $\times 10^{3}$ & $\times 10^{3}$ \\ \hline \hline
  $0^{\circ}$ & $1.55 $ & $0.206 $ \\ 
 $46^{\circ}$ & $1.57 $ & $0.205 $ \\ 
 $60^{\circ}$ & $1.54 $ & $0.212 $ \\ \hline 
\end{tabular}
\end{center}
\vspace{0.2cm}
{\it  Typical statistical errors on 
$< \omega >$ are of the order of $0.01 \times 10^{-3}$; those on $\sigma(\omega)$ are of the
order of $0.06 \times 10^{-4}$.}
\label{tab:zen}
\end{minipage}
\hspace{0.06\linewidth}
\begin{minipage}[t]{0.47\linewidth}
\caption{\it Average values and standard deviations of the $\omega$ distributions for
gamma-rays at zenith.}
\vspace{0.2cm}
\begin{center}
\begin{tabular}{ccc} \hline
Energy &  $< \omega >$ & $\sigma(\omega)$ \\ 
 (GeV) & $\times 10^{3}$ & $\times 10^{3}$ \\  \hline \hline
  $200$ & $1.516 $ & $0.301$ \\ 
 $500$ & $1.529 $ & $0.254 $ \\  
 $1000$ & $1.550 $ & $0.206 $ \\ 
 $5000$ & $1.650$ & $0.188 $ \\ \hline 
\end{tabular}
\end{center}
\vspace{0.2cm}
{\it Typical statistical errors on 
$< \omega >$ are of the order of $0.004 \times 10^{-3}$; those on $\sigma(\omega)$ are of the
order of $0.03 \times 10^{-4}$.}
\label{tab:en}
\end{minipage}
\end{table}

\section{Performance of the method}
\label{sec-perfo}
\subsection{The importance of the number of stereoscopic views}
\begin{figure}
\epsfig{file=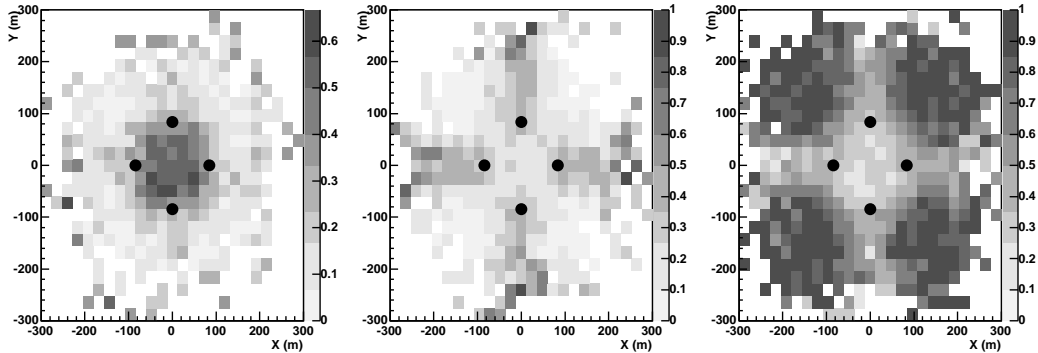,width=\linewidth}
\caption{\it Fraction of vertical $\gamma$-ray showers triggering $n_T$ telescopes 
as a function of the impact position (X and Y coordinates) on the ground. 
From left to right: $n_T=4$, $n_T=3$, $n_T=2$. The $\gamma$-ray spectrum is a power 
law with a spectral index
of 2.4. The black dots indicate the position of the H.E.S.S. telescopes.}
\label{fig:234tel}
\end{figure}
The performance of the method in terms of angular or spectral resolution depends on the
number $n_T$ of telescopes triggered by a shower; this number, referred to as
``telescope multiplicity'' in the following, depends in turn on the configuration
of the stereoscopic system and on the position of the shower core on the ground.
For the H.E.S.S. experiment, this is illustrated in figure \ref{fig:234tel} showing 
the distributions on the ground of the impacts of vertical $\gamma$-ray showers with 
$n_T=2$, $n_T=3$ and $n_T=4$ respectively;
in this example, $\gamma$-rays were simulated with a power-law spectrum of differential
spectral index 2.4. Events triggering 4 telescopes are concentrated in the central region of
the array, whereas events triggering 2 telescopes only are peripheral and extend further away
from the centre at higher energies. Of course, such peripheral events are not so accurately
reconstructed as the central ones; therefore, we shall investigate how the angular and
spectral resolution are modified when a minimal value of $n_T$ is required. It should be
noted that the improvement obtained with $n_T=4$ is not only due to the
redundancy of the reconstruction, but also to the relative position of the shower core 
with respect to the array.
\subsection{Gamma-ray selection efficiency (shower shape)}
\label{sec-eff}
\begin{figure}
\epsfig{file=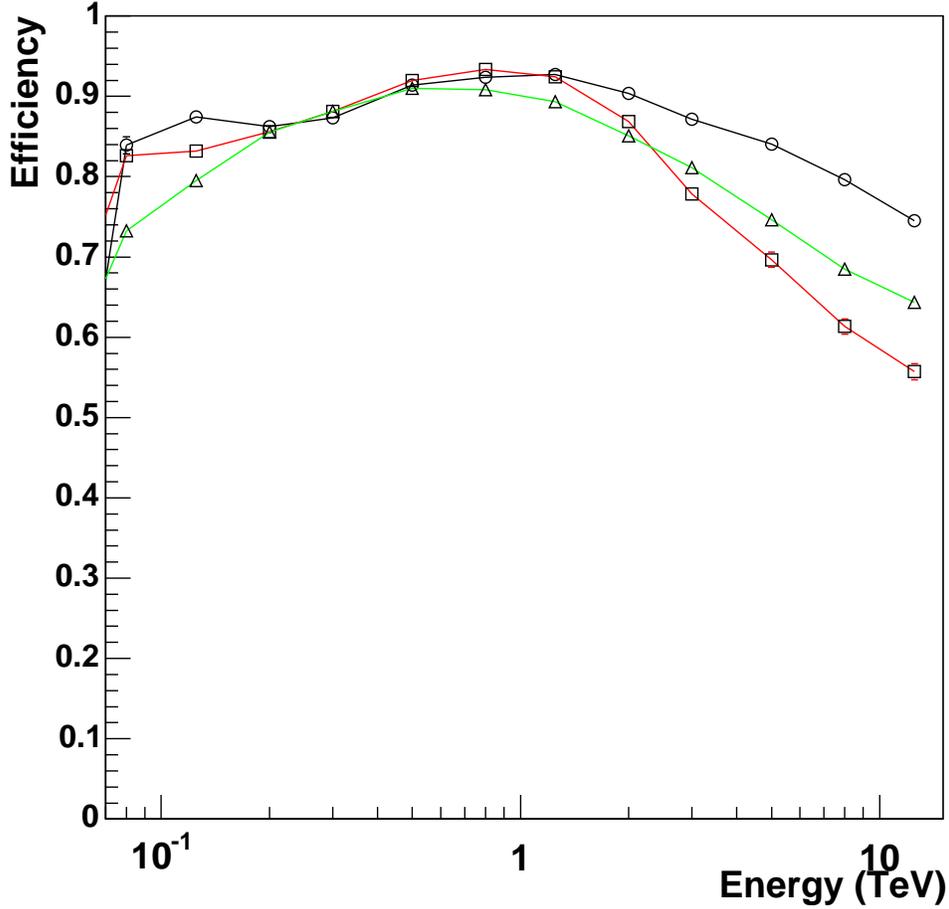,width=\linewidth}
\caption{\it Reconstruction efficiency for gamma-rays at zenith, as a function
of the primary energy, for showers triggering 2 telescopes (triangles), 3
telescopes (squares) and 4 telescopes (circles) respectively.}
\label{fig:effall}
\end{figure}
\begin{figure}
\begin{minipage}[t]{0.47\linewidth}
\epsfig{file=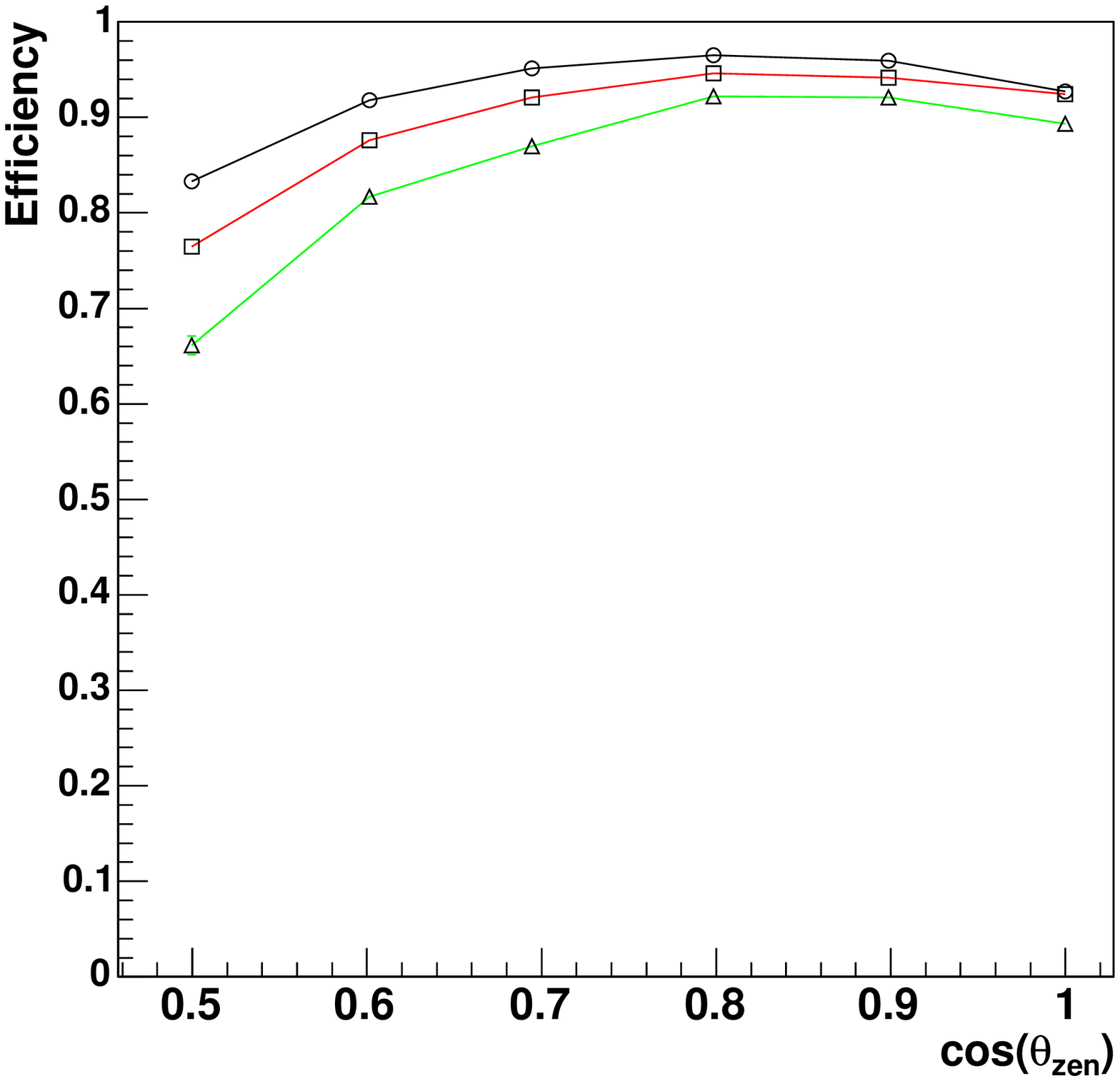,width=\linewidth}
\caption{\it  Reconstruction efficiency for 1~TeV~gamma-rays on axis as a function
of the zenith angle, for showers triggering 2~telescopes (triangles), 
3~telescopes (squares) and 4~telescopes (circles) respectively.}
\label{fig:effzen}
\end{minipage}
\hspace{0.06\linewidth}
\begin{minipage}[t]{0.47\linewidth}
\epsfig{file=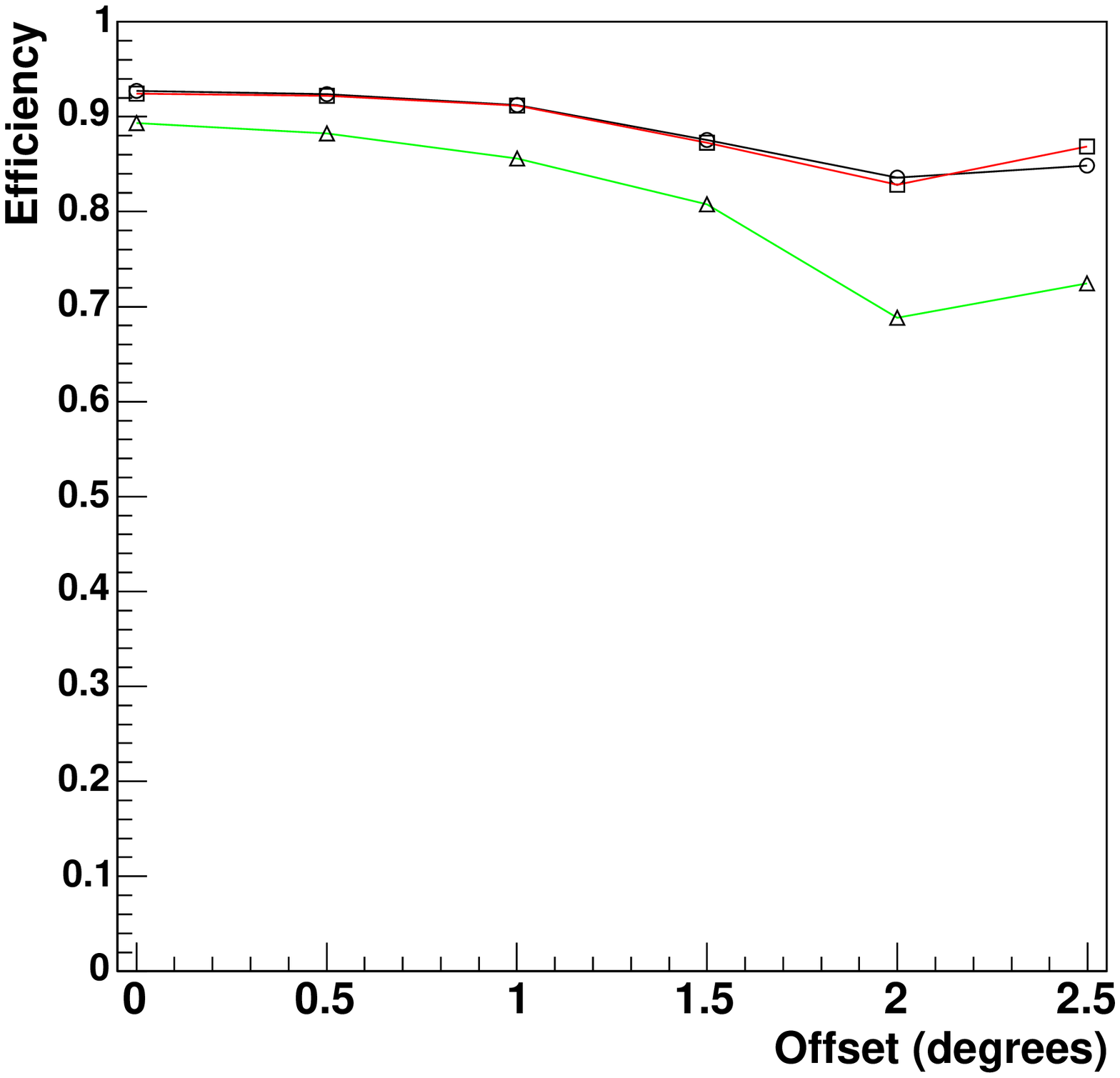,width=\linewidth}
\caption{\it  Reconstruction efficiency for 1~TeV~gamma-rays at zenith as a function
of the offset angle, for showers triggering 2~telescopes (triangles), 
3~telescopes (squares) and 4~telescopes (circles) respectively.}
\label{fig:effoff}
\end{minipage}
\end{figure}
The similarity between the distributions of $\omega$ at different zenith angles
and energies allows to select gamma-rays by using the same criterium
(\ref{eq:widthint}) on $\omega$ for all conditions of observation. 
On the basis of gamma-ray simulations at different zenith angles and energies, the
selection efficiency $\varepsilon_s$ for gamma-rays is defined as the fraction of events
accepted by the fit and satisfying  conditions (\ref{eq:physreg}) and 
(\ref{eq:widthint}). At this level, the selection
is only based on the shower shape, not on its direction, as is relevant 
in the study of extended sources\footnote{The global selection efficiency 
including the directional information (relevant for a point-like source) is
discussed later in section \ref{sec-hadrej}.}. Figure \ref{fig:effall} shows the variation of
$\varepsilon_s$ at zenith as a function of the primary energy for showers
triggering 2, 3 and 4 telescopes. Similarly, the variation of $\varepsilon_s$
as a function of the zenith angle is shown in figure \ref{fig:effzen} 
for 1~TeV $\gamma$-rays observed on axis; finally, for 1~TeV $\gamma$-rays observed 
off axis at zenith, figure \ref{fig:effoff} shows the variation of $\varepsilon_s$ 
with the offset angle, almost negligible for showers triggering 3 or 4 telescopes.
The relatively high value of $\varepsilon_s$ 
($>80$\% between 100~GeV and 3~TeV at zenith) and its smooth variation over 
a large range of energies and zenith angles are particularly well suited for
spectral analysis. 
\subsection{Angular resolution}
\label{sec-angres}
\begin{figure}
\begin{minipage}[t]{0.47\linewidth}
\epsfig{file=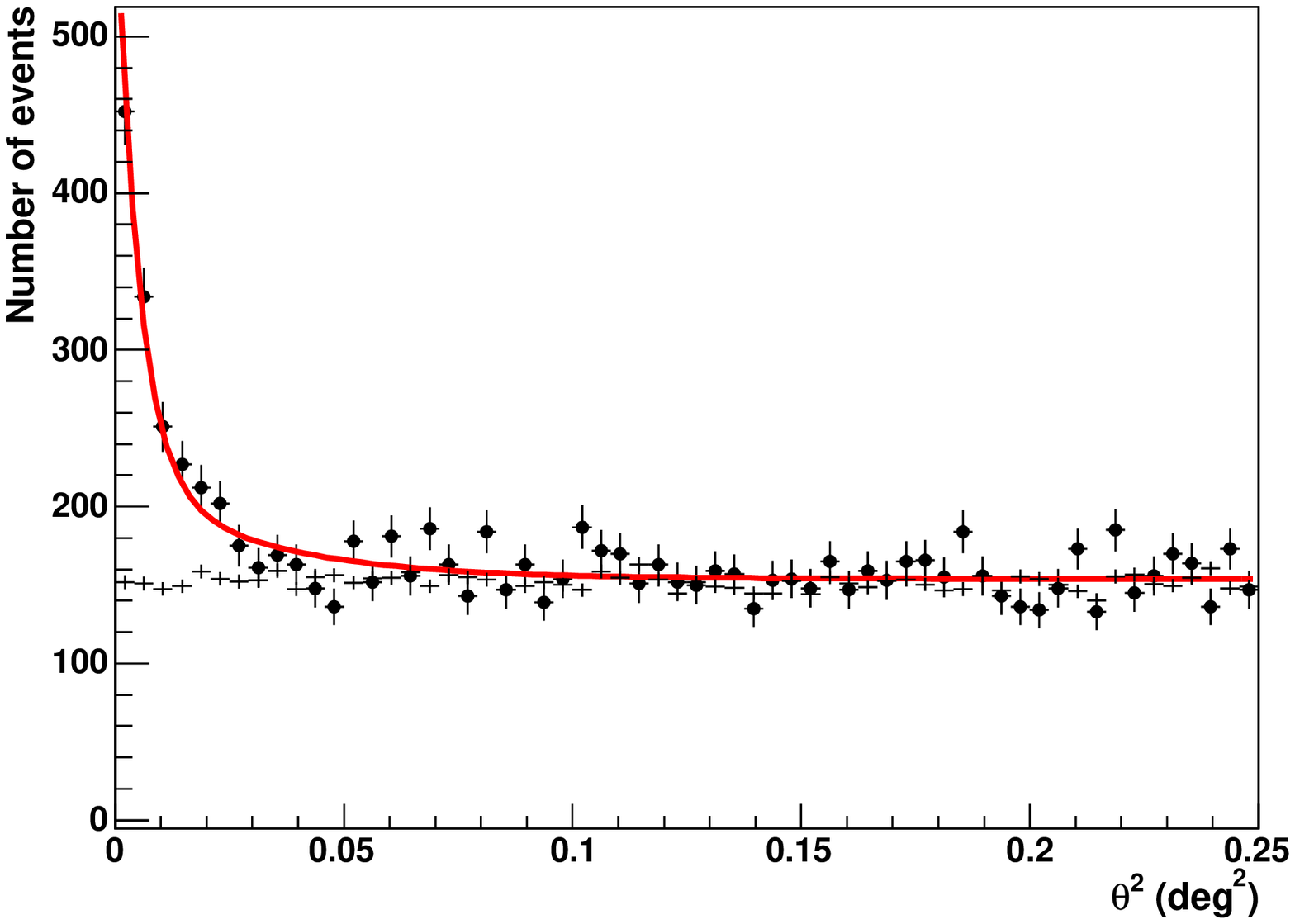,width=\linewidth}
\caption{\it $\theta^2$ distribution from data taken on PKS2155-304 in 2004:
(a) Filled circles: ``on source'' region. (b) Crosses: background control
regions.(c) The curve is the point-spread function obtained from simulated $\gamma$-rays with 
a  power-law spectrum of differential spectral index 3.2 at 26$^\circ$.}
\label{fig:pkangtot}
\end{minipage}
\hspace{0.06\linewidth}
\begin{minipage}[t]{0.47\linewidth}
\epsfig{file=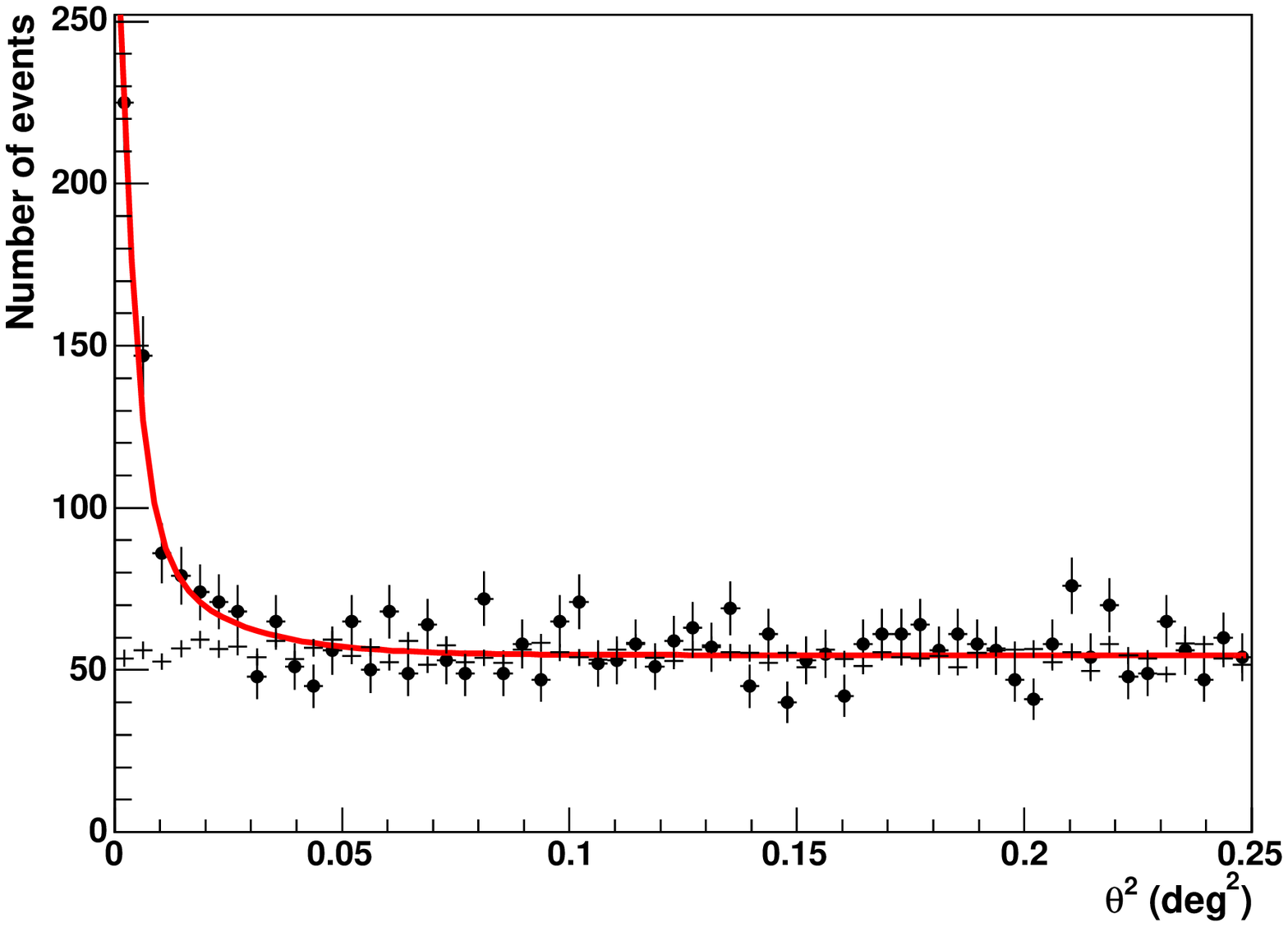,width=\linewidth}
\caption{\it $\theta^2$ distribution from data taken on PKS2155-304 in 2004,
restricted to showers triggering at least 3 telescopes:
(a) Filled circles: ``on source'' region. (b) Crosses: background control
regions. The curve is the point-spread function obtained from simulated $\gamma$-rays with 
a  power-law spectrum of differential spectral index 3.2 at 26$^\circ$, keeping only events 
triggering at least 3 telescopes.}
\label{fig:pkang34t}
\end{minipage}
\end{figure}

\begin{figure}
\epsfig{file=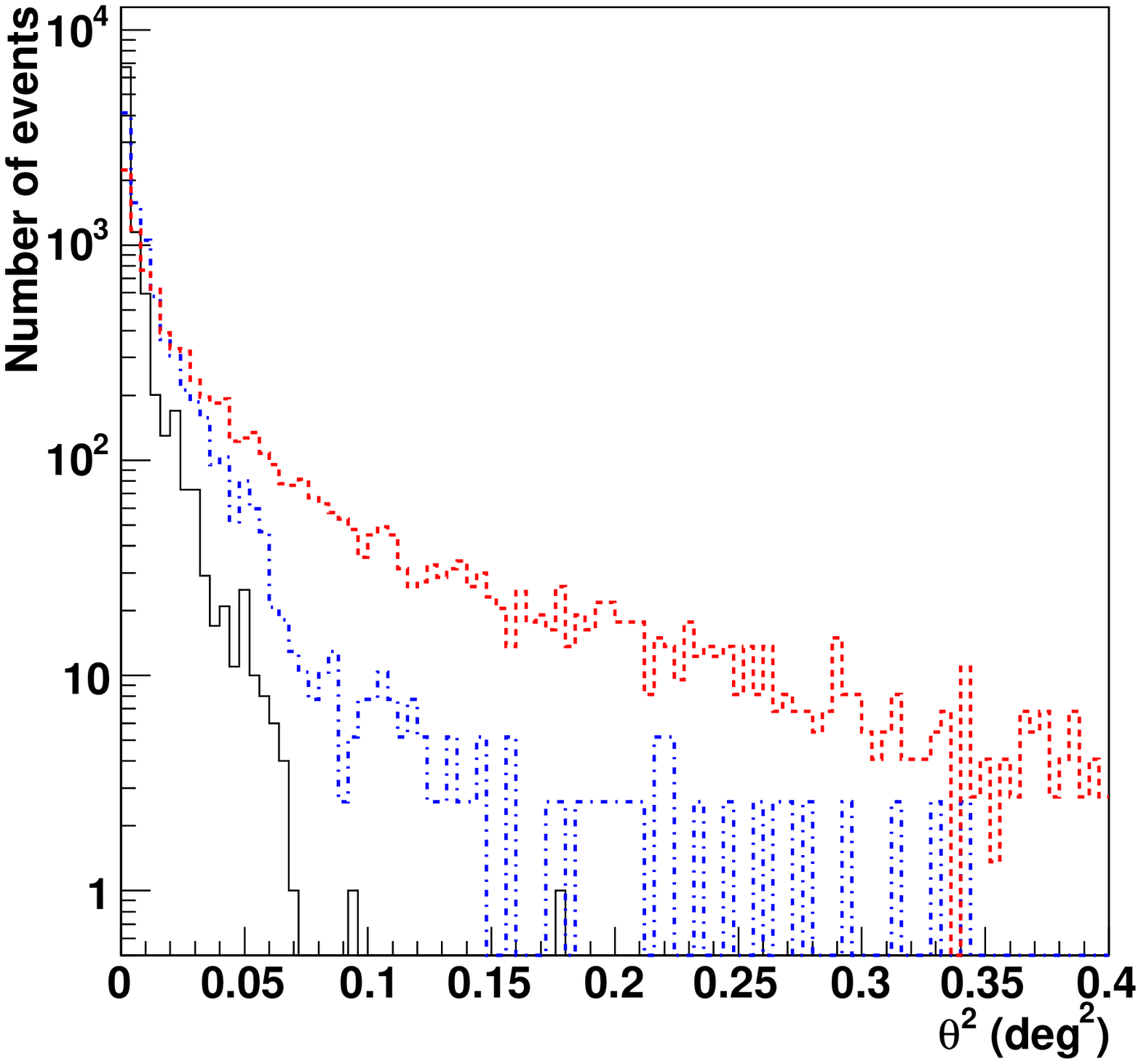,width=\linewidth}
\caption{\it $\theta^2$ distribution from 1~TeV $\gamma$-ray showers simulated
at zenith. (a) Dashed line: events triggering two telescopes. (b) Dashed-dotted line:
events triggering three telescopes. (c) Solid line: events triggering four telescopes.}
\label{fig:angres}
\end{figure}

\begin{figure}
\begin{minipage}[t]{0.47\linewidth}
\epsfig{file=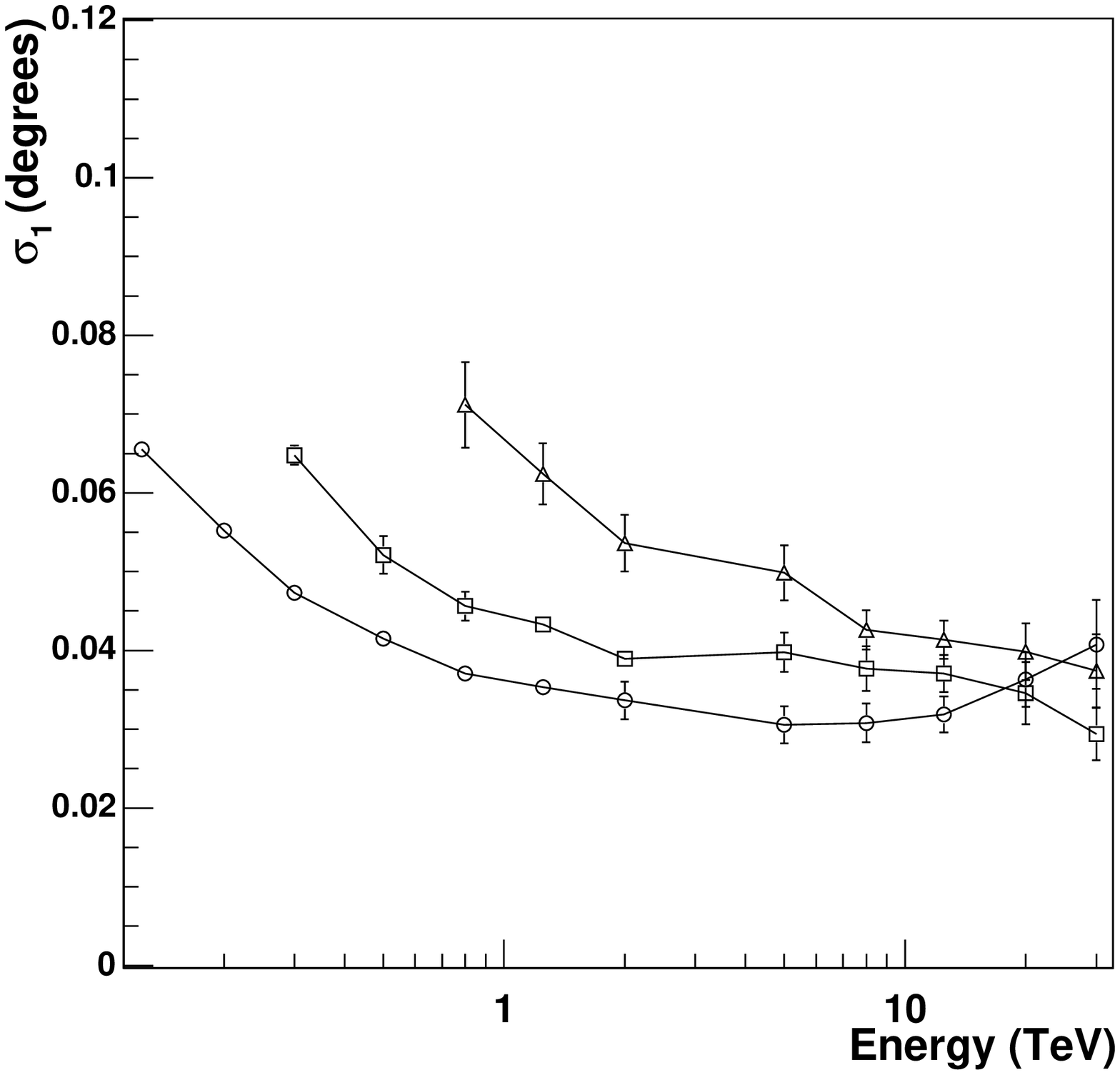,width=\linewidth}
\caption{\it Central spot spread $\sigma_1$, as a function of $\gamma$-ray energy for zenith
angles 0$^{\circ}$ (circles), 46$^{\circ}$ (squares) and 60$^{\circ}$
(triangles).}
\label{fig:spot}
\end{minipage}
\hspace{0.06\linewidth}
\begin{minipage}[t]{0.47\linewidth}
\epsfig{file=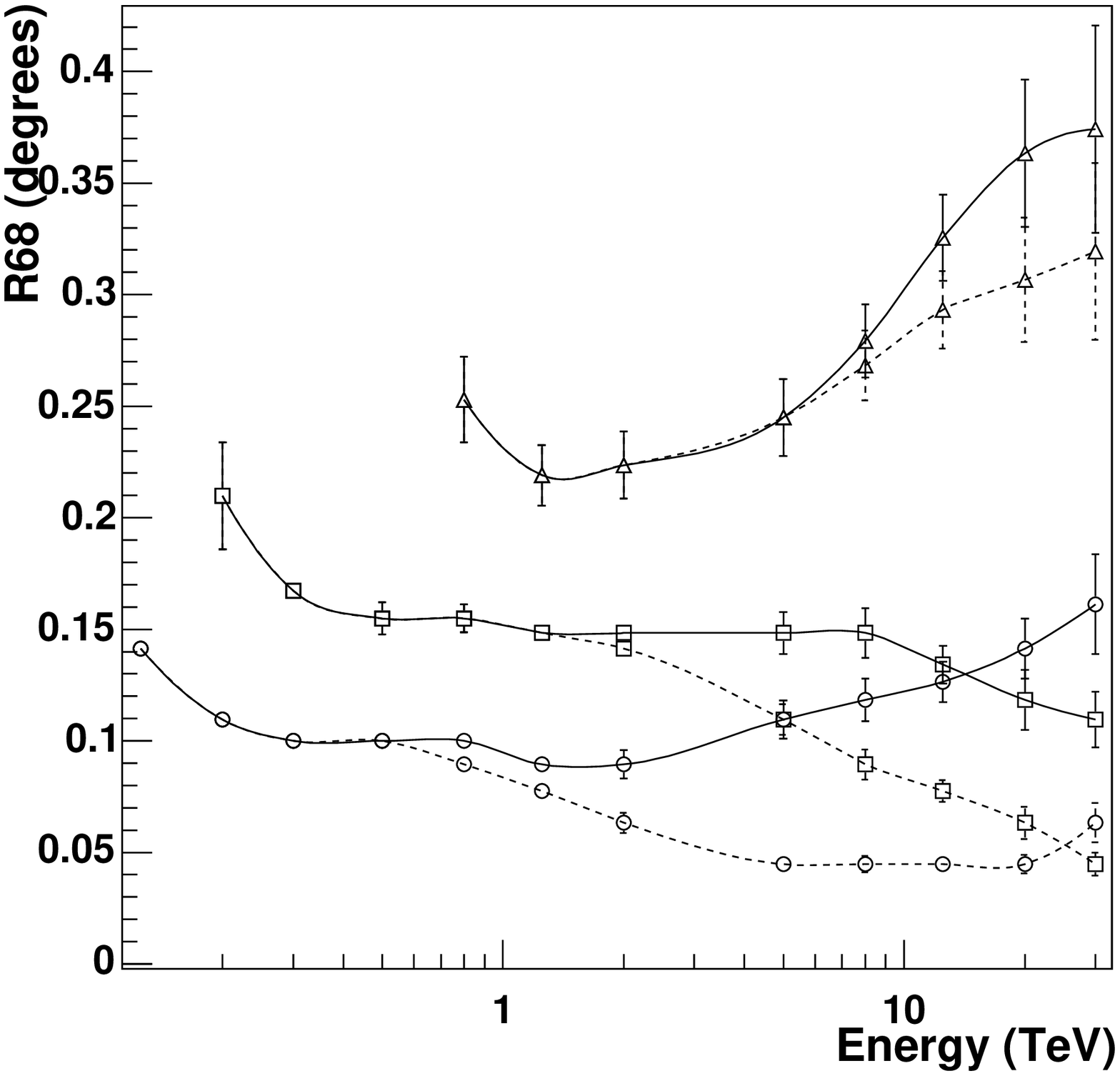,width=\linewidth}
\caption{\it Radius of the 68\% containment cone,as a function of $\gamma$-ray energy for zenith
angles 0$^{\circ}$ (circles), 46$^{\circ}$ (squares) and 60$^{\circ}$
(triangles). Dotted lines show the effect of an additional cut
requiring at least two images whose centre of gravity is 
within 2$^{\circ}$ from the camera centre.}
\label{fig:r68}
\end{minipage}
\end{figure}

\begin{figure}
\begin{minipage}[t]{0.47\linewidth}
\epsfig{file=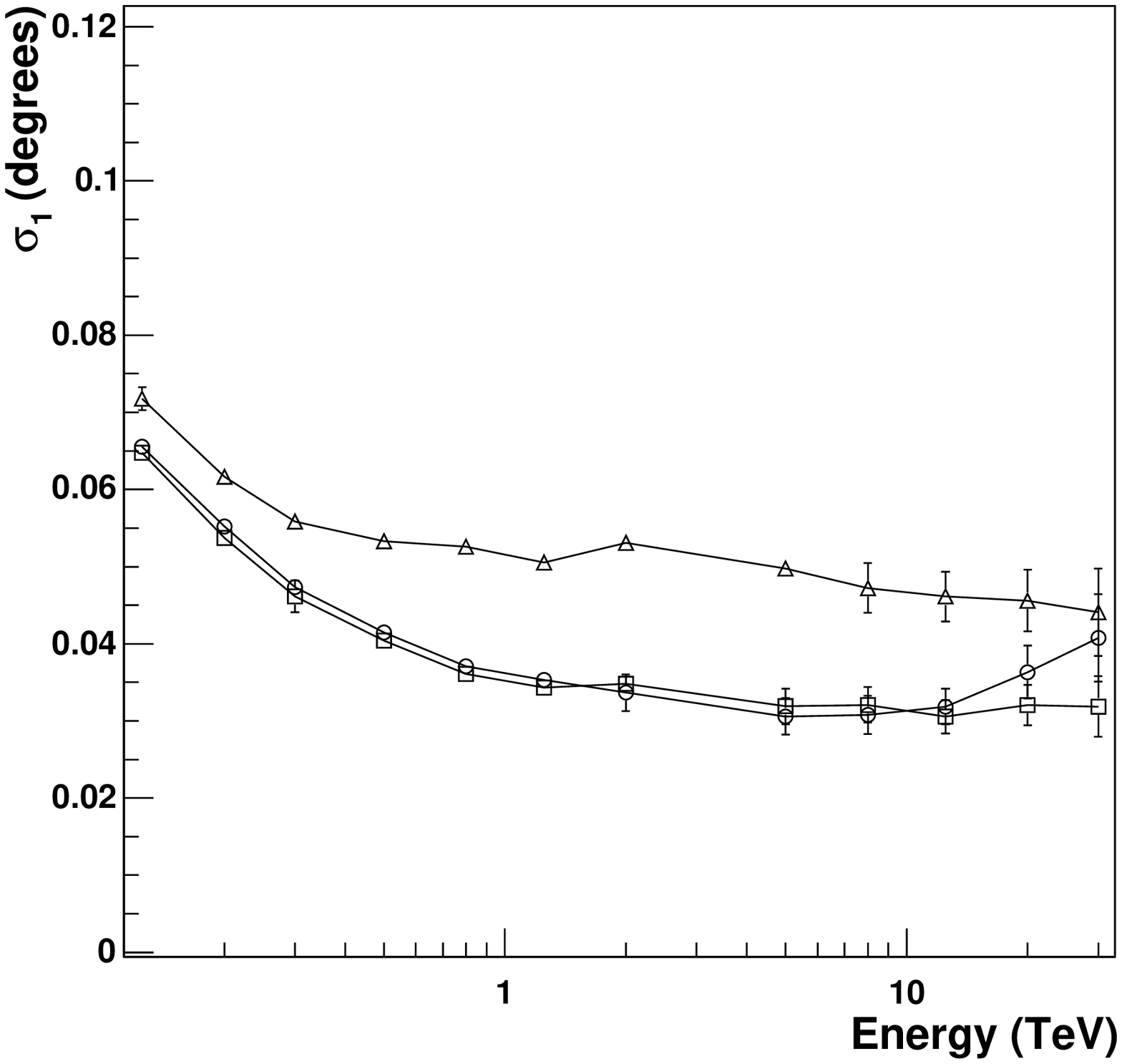,width=\linewidth}
\caption{\it Central spot spread $\sigma_1$, as a function of $\gamma$-ray energy for offset
angles 0$^{\circ}$ (circles), 1$^{\circ}$ (squares) and 2$^{\circ}$ (triangles); 
$\gamma$-rays are simulated at zenith.}
\label{fig:spotoff}
\end{minipage}
\hspace{0.06\linewidth}
\begin{minipage}[t]{0.47\linewidth}
\epsfig{file=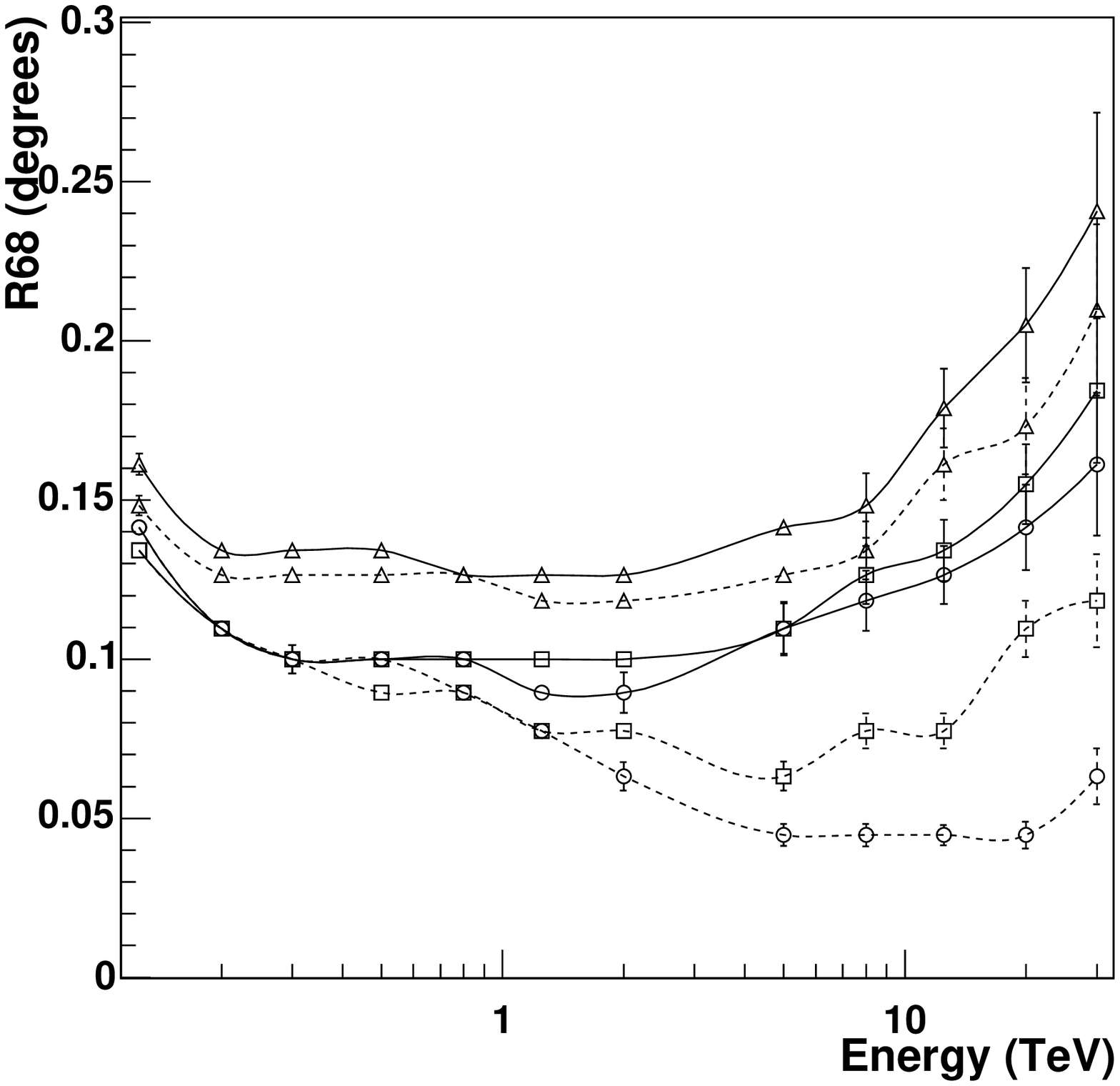,width=\linewidth}
\caption{\it Radius of the 68\% confidence level region,as a function of $\gamma$-ray energy for offset
angles 0$^{\circ}$ (circles), 1$^{\circ}$ (squares) and 2$^{\circ}$
(triangles); $\gamma$-rays are simulated at zenith. 
Dotted lines show the effect of an additional cut
requiring at least two images whose centre of gravity is 
within 2$^{\circ}$ from the camera centre.}
\label{fig:r68off}
\end{minipage}
\end{figure}

The angular resolution is characterized by the ``point-spread function'' (PSF),
i.e. the distribution of the angle $\theta$ between the reconstructed direction and that
of the source. It is obtained from data taken on a distant extragalactic source,
e.g. PKS2155-304. For the pupose of background monitoring, the source was observed at 
an angular distance $\alpha = 0.5^{\circ}$ from the telescope axis (offset angle).
Figure \ref{fig:pkangtot} shows the $\theta^2$ distribution
from the \hess~data taken on this source and already referred to in 
section \ref{sec-3Dw}. The angular
resolution is further improved by restricting to showers triggering at least 3
telescopes (figure \ref{fig:pkang34t}). The PSF obtained from simulations in a configuration 
close to the one of PKS2155-304 ($\gamma$-ray 
power-law spectrum  with a differential spectral index of 3.2 at 26$^\circ$ zenith angle) 
was superimposed on the histograms, showing the good agreement between the $\theta^2$ 
distributions obtained with real data and those from simulations.
Details of the $\theta^2$ distribution are more clearly visible in figure
\ref{fig:angres} in which the vertical scale is logarithmic; this distribution
is obtained from 1 TeV gamma-ray showers simulated at zenith and further
reconstructed as explained above. It is well fitted
by a linear superposition of two exponential laws in $\theta^2$, 
\[ \frac{dP}{d \theta^2} = \frac{\alpha}{2 \sigma_1^2} 
\exp \left(-\frac{\theta^2}{2 \sigma_1^2}\right) + 
\frac{1-\alpha}{2 \sigma_2^2} \exp \left(-\frac{\theta^2}{2 \sigma_2^2}\right) \]
i.e. of two Gaussian laws in projected angles ($\theta_x$ or $\theta_y$); 
the narrower one, referred to as ``the central spot'' and characterized by $\sigma_1$
is thus superimposed to a broader halo characterized by $\sigma_2$.
The halo is more important for those events which trigger 2~telescopes 
only (figure \ref{fig:angres} (a)); this effect, which is independent of the 
reconstruction method, is mainly related to the position of the impact of the 
shower axis on the ground, relatively far away from the centre of the array 
for those events (figure~\ref{fig:234tel}). 
On the other hand, the angular distribution is practically
reduced to the central spot for events triggering 4 telescopes
(figure \ref{fig:angres} (c)).
Figure \ref{fig:spot}, also obtained from simulations, shows the variation of 
the spread of the central spot $\sigma_1$ as a function of the $\gamma$-ray energy for 
different zenith angles and on-axis showers; events with all telescope multiplicities are included.
This spread is practically always smaller than 0.06$^{\circ}$ (about 4$^{\prime}$) and
remains fairly constant at energies greater than 1 TeV. In order to take the
halo into account, one can also characterize the angular distribution by the 
angular radius $R_{68}$ of the cone centered on the true $\gamma$-ray direction and
containing 68\% of the reconstructed axes. Its variation as a function of 
the $\gamma$-ray energy for different zenith angles is shown in figure
\ref{fig:r68} in which events with all telescope multiplicities are included.
In the case of off-axis observations, the influence of the offset angle $\alpha$ on the
angular resolution is shown in figure \ref{fig:spotoff} for $\sigma_1$ and 
in figure \ref{fig:r68off} for $R_{68}$; it is practically negligible for 
$\alpha < 1^{\circ}$. By comparing figures \ref{fig:spot} and \ref{fig:r68},
it appears that the relative importance of the halo strongly depends
on the zenith angle and on the position of the impact of the
shower axis; in particular, the increase of $R_{68}$ with energy above a
few TeV is due to the contribution of energetic showers triggering the array 
from rather remote impacts ($\sim 300~m$ from the centre). This effect is
attenuated if one requires at least two images whose centre of gravity is 
within 2$^{\circ}$ from the camera centre, a cut used in the traditional 
analysis based on Hillas parameters (dashed lines in figure \ref{fig:r68}).

\begin{figure}
\begin{minipage}[t]{0.47\linewidth}
\epsfig{file=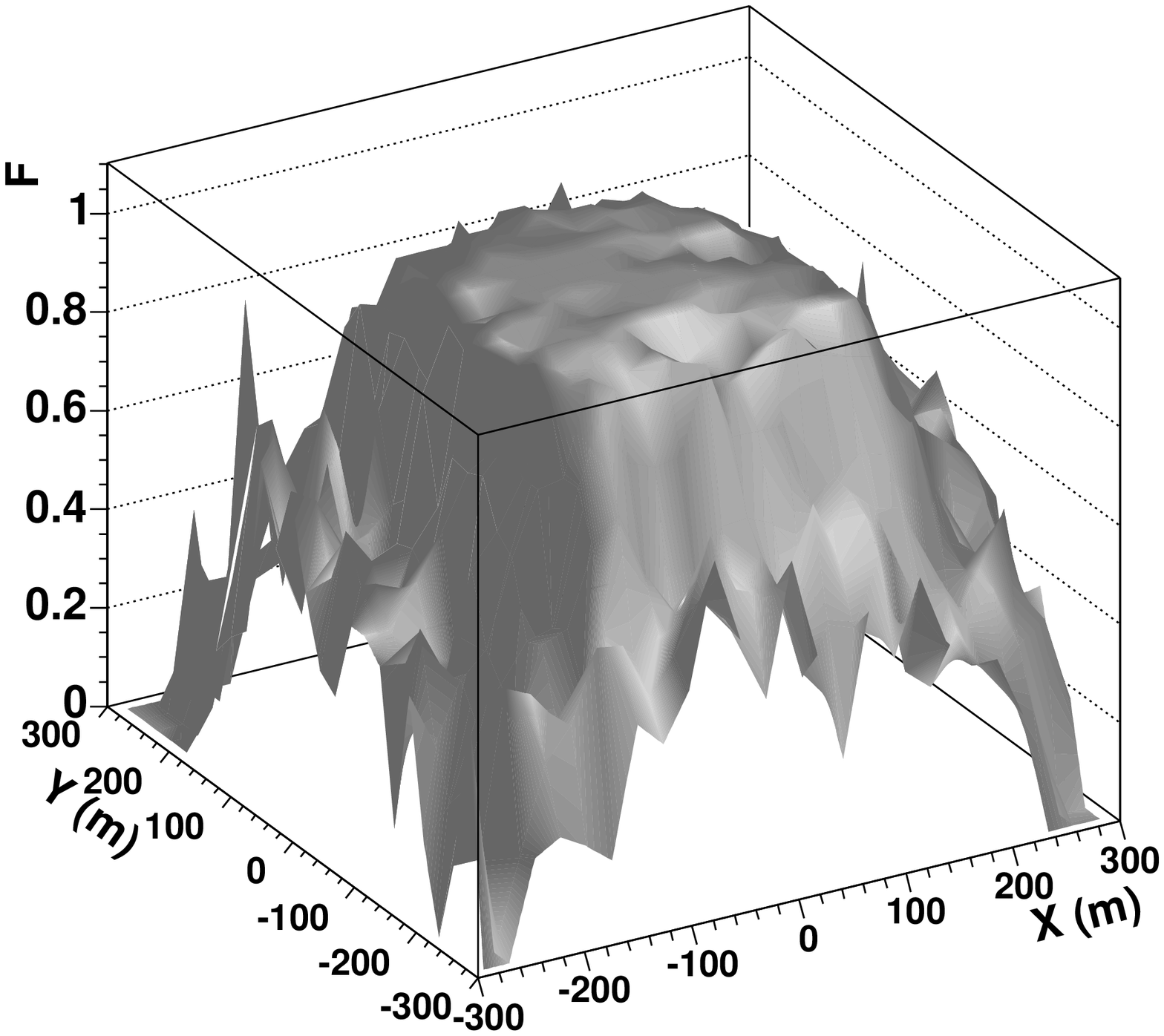,width=\linewidth}
\caption{\it Fraction $F$ of 1~TeV vertical $\gamma$-ray showers with
$\theta < 0.1^{\circ}$ as a function of the impact position 
($X$ and $Y$ coordinates in $m$) on the ground.}
\label{fig:goodangres}
\end{minipage}
\hspace{0.06\linewidth}
\begin{minipage}[t]{0.47\linewidth}
\epsfig{file=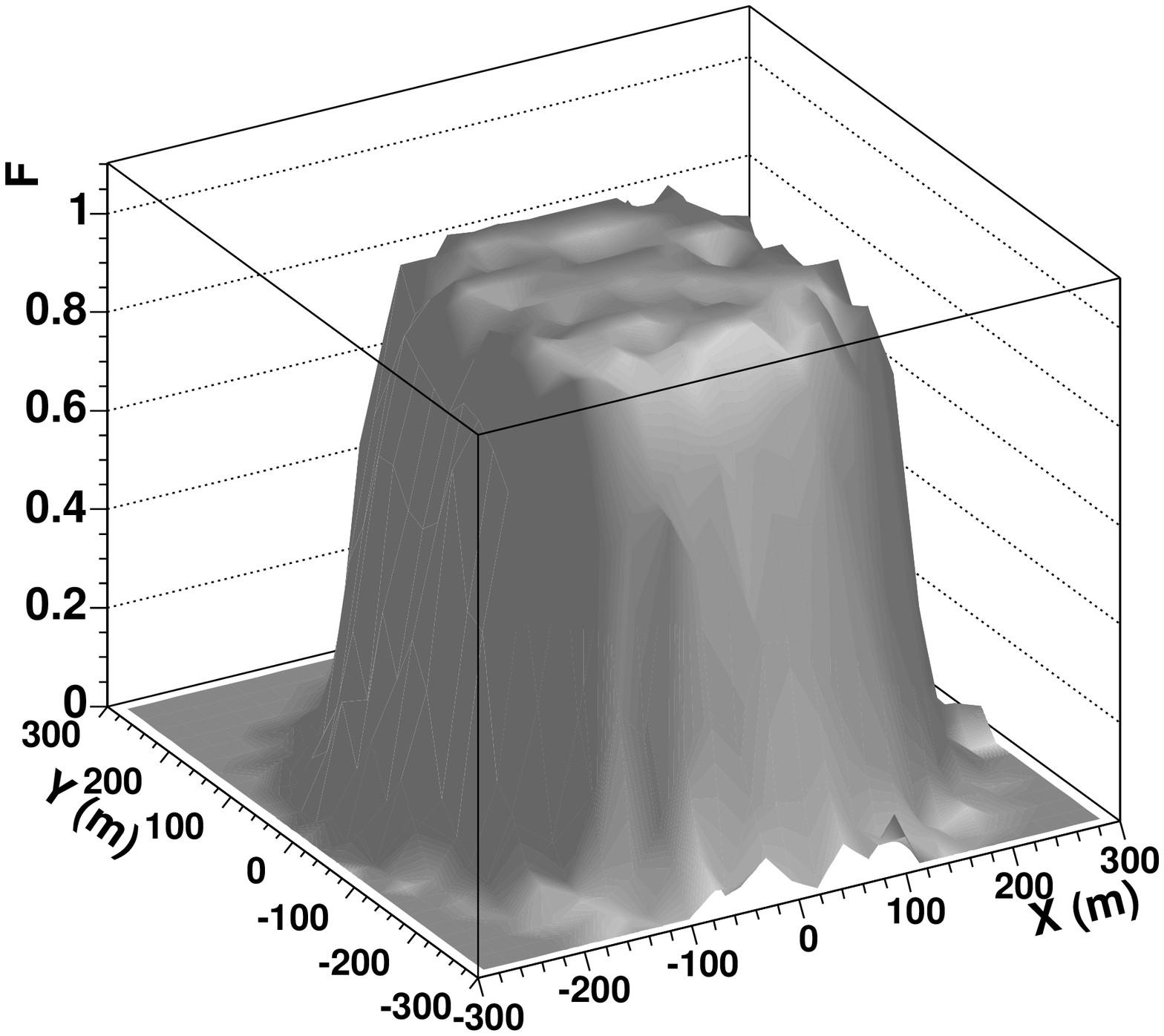,width=\linewidth}
\caption{\it Fraction $F$ of 1~TeV vertical $\gamma$-ray showers triggering 4
telescopes as a function of the impact position ($X$ and $Y$ coordinates in $m$)
on the ground.}
\label{fig:core4tels}
\end{minipage}
\end{figure}

The influence on the angular resolution of the position of the shower impact 
on the ground is shown in figures \ref{fig:goodangres} and \ref{fig:core4tels}, 
obtained from simulations of vertical 1~TeV $\gamma$-ray showers observed on-axis. 
Figure \ref{fig:goodangres} refers to well-reconstructed showers, i.e. to those events 
whose reconstructed direction
differs from the true one by less than 0.1$^{\circ}$; the figure shows the
fraction of this population among the detected events as a function of the impact
position. This fraction is remarkably uniform --- around 80\% --- in a large 
250~m~$\times$~250~m square area, and is still quite significant (about 40\%)
300~m away from the centre of the array. The extension of well-reconstructed events
is thus significantly larger than that of showers with $n_T=4$ shown in figure
\ref{fig:core4tels} for comparison.
\subsection{Global $\gamma$-ray selection efficiency and hadronic rejection}
\label{sec-hadrej}
\begin{figure}
\epsfig{file=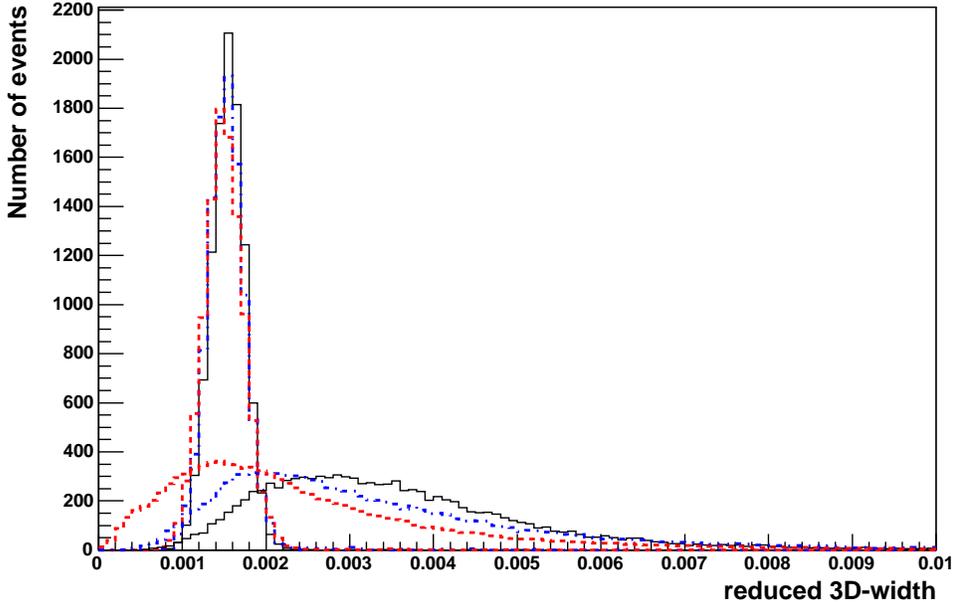,width=\linewidth}
\caption{\it Distributions of the reduced 3D-width $\omega$ for $\gamma$-rays (peaks) and for hadrons for
different observation conditions. (a) Dashed line: events triggering two telescopes. (b) Dashed-dotted line:
events triggering three telescopes. (c) Solid line: events triggering four telescopes. Gamma-rays
were simulated at 1~TeV and a zenith angle of 18$^{\circ}$; hadrons were taken from a field of
view free of $\gamma$-ray sources at a zenith angle of 16$^{\circ}$. All histograms are normalized to the same
number of events.}
\label{fig:compgh}
\end{figure}

The selection criteria used in section \ref{sec-eff} were based exclusively on the shower shape. 
The corresponding efficiencies $\varepsilon_s$, averaged over a $\gamma$-ray power-law spectrum with 
a differential spectral index of 2.4, are shown in table \ref{tab:rejeff} for different zenith angles 
and for different requirements on the minimal telescope multiplicity $n_T$. The same criteria
reduce the number of hadrons by a factor $R_s$ (rejection factor); the values of $R_s$ shown in table 
\ref{tab:rejeff} were obtained from real data taken in fields of view free of $\gamma$-ray sources.
Whereas $\varepsilon_s$ is rather insensitive to $n_T$, a significant improvement of the hadronic 
rejection factor $R_s$ is obtained when requiring a minimal telescope multiplicity of 3.
This is essentially due to the cut in the reduced 3D-width $\omega$, since the distribution of this variable
for hadrons strongly depends on $n_T$ as shown in figure \ref{fig:compgh}; clearly, the
constraint of rotational symmetry is more accurately checked for those showers
surrounded by 3 or 4 telescopes. On the other hand, neither $\varepsilon_s$, nor $R_s$ vary
much with the zenith angle $\zeta$.
\begin{table}[htbp]
\caption{\it Gamma-ray selection efficiencies $\varepsilon_s$ (shape criteria) and  $\varepsilon_g$
(shape + direction) and hadronic rejection factors corresponding to the same criteria, $R_s$ and $R_g$
respectively, for different zenith angles and for different requirements on the minimal telescope multiplicity $n_T$.
Gamma-rays were simulated with a power-law spectrum with a differential spectral index of
2.4. Hadron rejection factors were obtained from real data taken in fields of view free of $\gamma$-ray
sources. Global quality factors $Q_g$ (see text and footnote for the definition)
are also shown in the table.} 
\vspace{0.2cm}
\begin{center}
\begin{tabular}{|c||ccc|ccc|ccc|} \hline
Zenith Angle & \multicolumn{3}{|c|}{18$^{\circ}$} & \multicolumn{3}{|c|}{33$^{\circ}$} &
\multicolumn{3}{|c|}{51$^{\circ}$} \\ \hline \hline
$n_T$ & $\ge 2$ & $\ge 3$ & $ = 4$ & $\ge 2$ & $\ge 3$ & $ = 4$ & $\ge 2$ & $\ge 3$ & $ = 4$ \\ \hline
$\varepsilon_s$ & 0.86 & 0.90 & 0.90 & 0.89 & 0.92 & 0.93 & 0.85 & 0.89 & 0.90 \\
$\varepsilon_g$ & 0.51 & 0.67 & 0.77 & 0.55 & 0.70 & 0.80 & 0.53 & 0.63 & 0.70 \\ \hline \hline
$R_s$ & 8.9 &  16.1 &  22.4 & 8.7 & 13.9 & 18.3 & 8.3 & 11.3 & 13.9 \\
$R_g$ & 5000 & 6870 & 7670 & 4070 & 4980 & 4750 & 2010 & 2170 & 2880 \\ \hline \hline
$Q_g$ & 36 & 56 & 67 & 35 & 49 & 55 & 24 & 30 & 38 \\ \hline \hline
\end{tabular}
\end{center}
\label{tab:rejeff}	   	       
\end{table}	

In the study of a point-like source, the $\gamma$-ray selection criteria based on shower shape must be 
complemented by a cut on the angle $\theta$ between the reconstructed direction and that of the source. 
For the sake of the present example, we require $\theta$ to be lower than $0.1^{\circ}/\cos \zeta$; this
choice reflects the variation of the angular resolution with the zenith angle $\zeta$.
The global $\gamma$-ray selection efficiencies $\varepsilon_g$ and the global hadronic
rejection factor $R_g$, obtained after this last cut, are shown in table \ref{tab:rejeff} for different 
observation conditions. The global selection efficiency $\varepsilon_g$ is almost independent of the zenith angle and
remains of the order of 50\% with no restriction on $n_T$ and reaches 75\% for $n_T=4$; as a matter of fact,
the angular cut removes most of the events from the halo (defined according to section \ref{sec-angres})
whose contribution is significant when no restriction on $n_T$ is applied and small for $n_T=4$. The global hadronic
rejection factor varies between 2000 and 5000 with no restriction on $n_T$ (and between 2800 and 8000 for
$n_T=4$). One often defines a quality factor $Q_g = \varepsilon_g \sqrt{R_g}$ to characterize the sensitivity 
of an analysis method\footnote{The quality factor thus defined is however not a good standard of comparison
between different experiments, since some fraction of background events is already removed at the trigger level.
In particular, muons, which make a significant part of the background events for a single telescope are
eliminated by the central trigger in a stereoscopic system.}. With the rather conservative cuts used above, 
$Q_g$ is of the order of 20 to 40 
with no restriction on $n_T$ and is practically doubled when one requires $n_T=4$. It should be noted that
a minimal set of cuts have been used in the preceding example and that no optimization procedure has been
applied. Additional requirements (e.g. on the minimum number of photo-electrons per telescope) and different
sets of cuts can be used to improve the analysis, depending on the intensity of the source.

\subsection{Gamma-ray energy measurement}
\label{sec-ener}
\begin{figure}
\begin{minipage}[t]{0.47\linewidth}
\epsfig{file=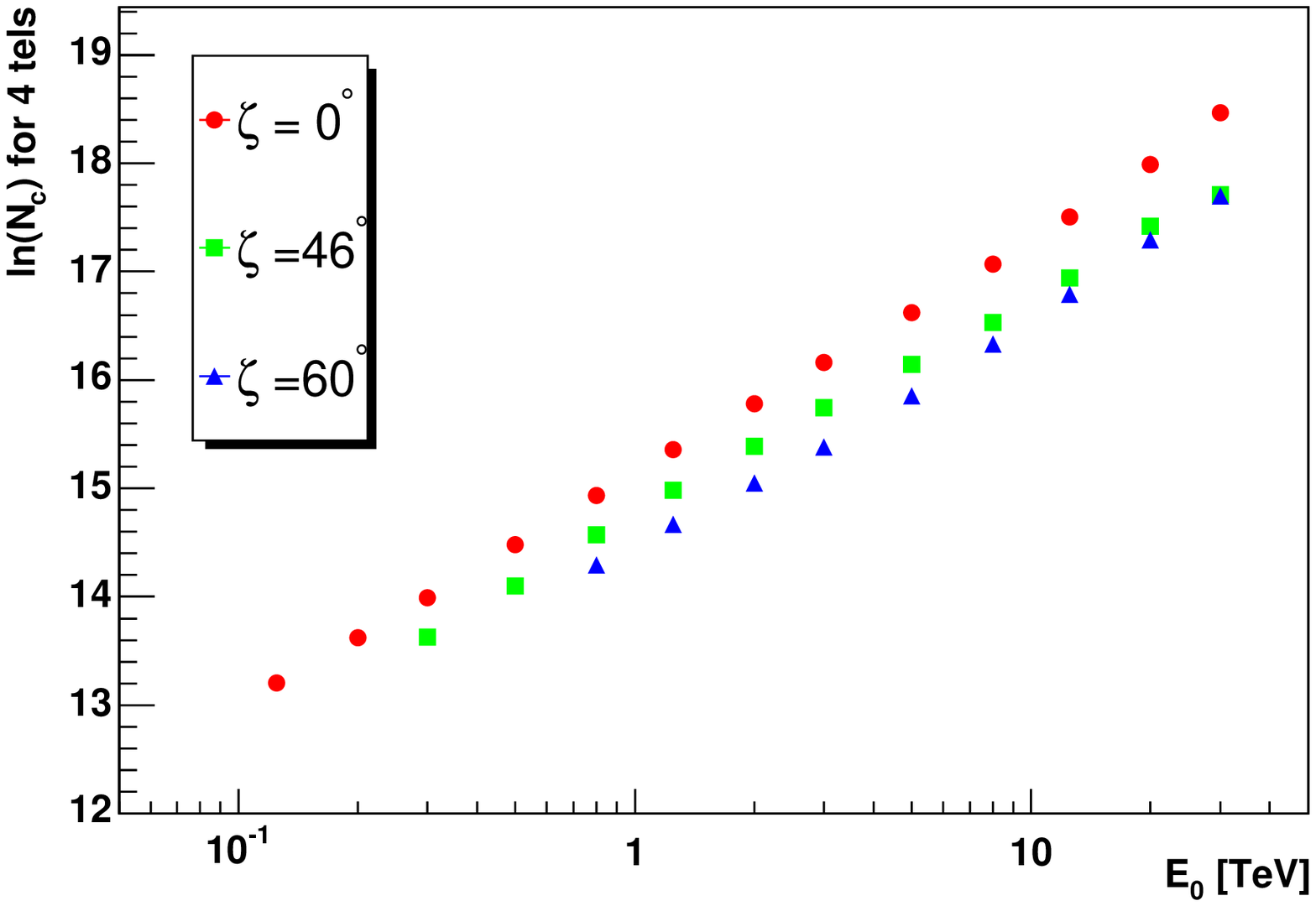,width=\linewidth}
\caption{\it Average value of $\ln N_c$ as a function of $E_0$ for $\gamma$-ray
showers simulated on axis at different zenith angles: $0^{\circ}$ (circles), 
$46^{\circ}$ (squares) and $60^{\circ}$ (triangles), and triggering 4 telescopes.}
\label{fig:calibzen}
\end{minipage}
\hspace{0.06\linewidth}
\begin{minipage}[t]{0.47\linewidth}
\epsfig{file=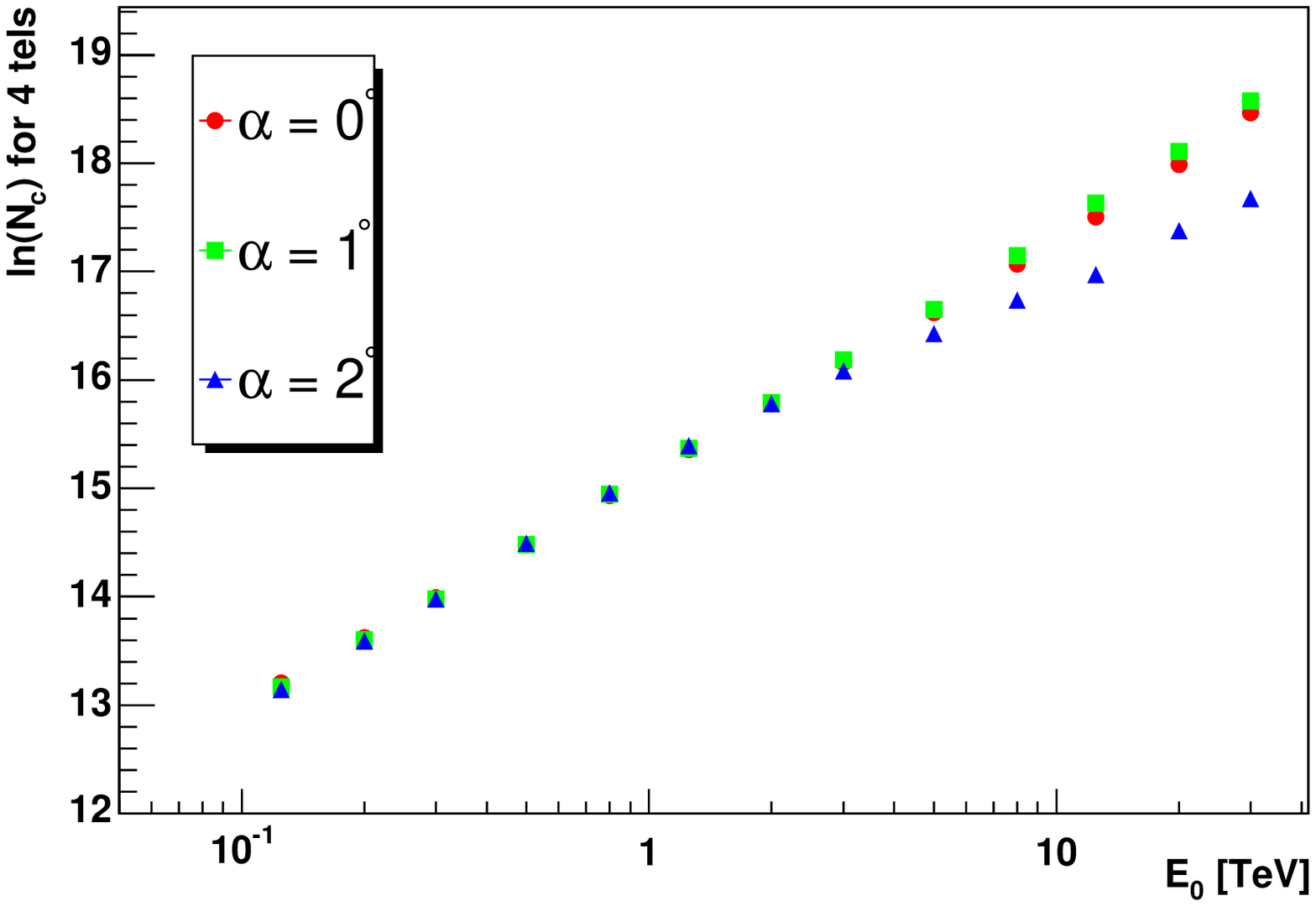,width=\linewidth}
\caption{\it Average value of $\ln N_c$ as a function of $E_0$ for $\gamma$-ray
showers simulated at zenith for two offset angles: $0^{\circ}$ (circles), $1^{\circ}$ (squares) 
and $2^{\circ}$ (triangles), and triggering 4 telescopes.}
\label{fig:caliboff}
\end{minipage}
\end{figure}
The energy $E_0$ of the primary gamma-ray
is reconstructed calorimetrically from the total number of Cherenkov photons 
$N_c$ obtained from the fit. If the 3D-model of the shower described above
were a perfect representation of 
the electromagnetic shower, $N_c$ would be, on average, almost\footnote{The
variation of the Cherenkov threshold as a function of the altitude makes this statement only approximate.} 
proportional to $E_0$. As a matter of fact, simulations
of $\gamma$-ray showers for fixed values of $E_0$, show that the average value 
of $\ln N_c$ (noted as $\big< \ln N_c \big>$), obtained from the likelihood fit, 
varies almost linearly with $\ln E_0$ 
{\it for fixed observing conditions}, namely: telescopes multiplicity $n_T$,
direction of the telescope axes (i.e. zenith angle $\zeta$), direction of the shower axis
(i.e. offset angle $\alpha$). For those showers triggering 4 telescopes,
i.e. those for which the sampling is more homogeneous, the slope 
$a = \partial \big< \ln N_c \big>/ \partial \ln E_0$
is actually close to 1; this is shown in figure \ref{fig:calibzen} for showers generated on-axis at
different zenith angles, as well as in figure
\ref{fig:caliboff} for showers at zenith generated off-axis at different angles.\\

\begin{figure}
\begin{minipage}[t]{0.47\linewidth}
\epsfig{file=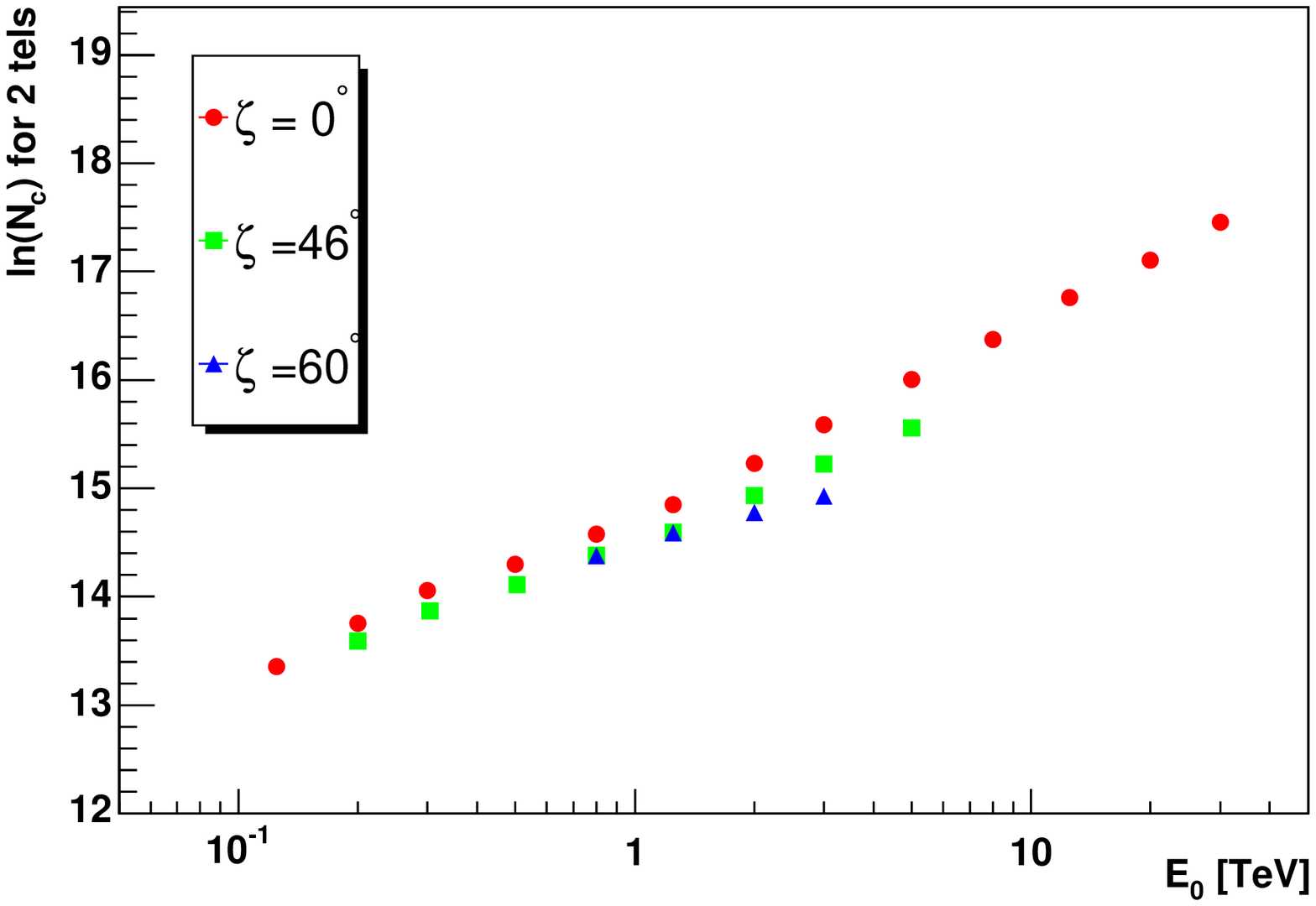,width=\linewidth}
\caption{\it Average value of $\ln N_c$ as a function of $E_0$ for $\gamma$-ray
showers simulated on axis at different zenith angles: $0^{\circ}$ (circles), 
$46^{\circ}$ (squares) and $60^{\circ}$ (triangles), and triggering 2 telescopes.}
\label{fig:calibzen2}
\end{minipage}
\hspace{0.06\linewidth}
\begin{minipage}[t]{0.47\linewidth}
\epsfig{file=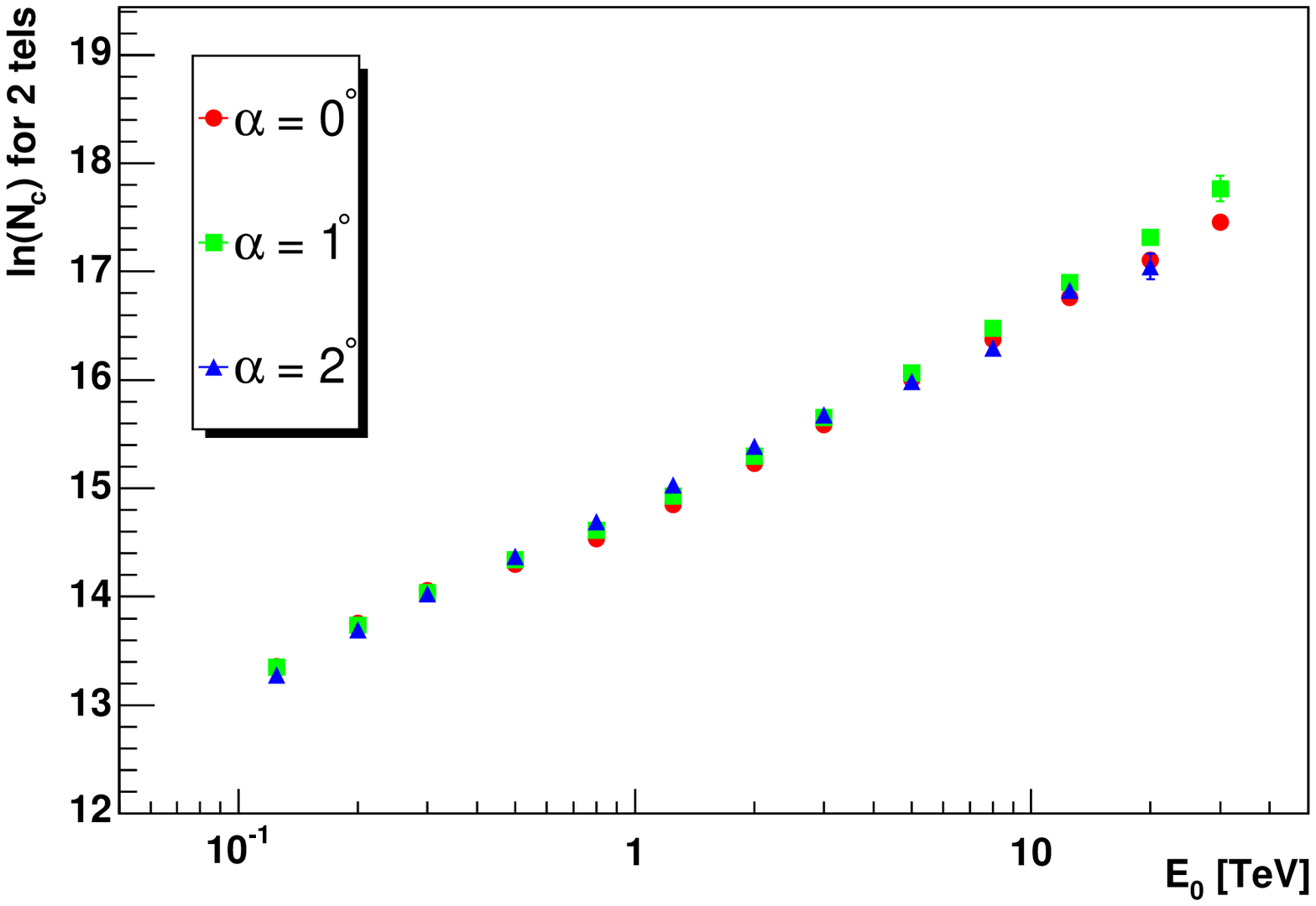,width=\linewidth}
\caption{\it Average value of $\ln N_c$ as a function of $E_0$ for $\gamma$-ray
showers simulated at zenith for two offset angles: $0^{\circ}$ (circles), $1^{\circ}$ (squares) 
and $2^{\circ}$ (triangles), and triggering 2 telescopes.}
\label{fig:caliboff2}
\end{minipage}
\end{figure}

\begin{figure}
\begin{minipage}[t]{0.47\linewidth}
\epsfig{file=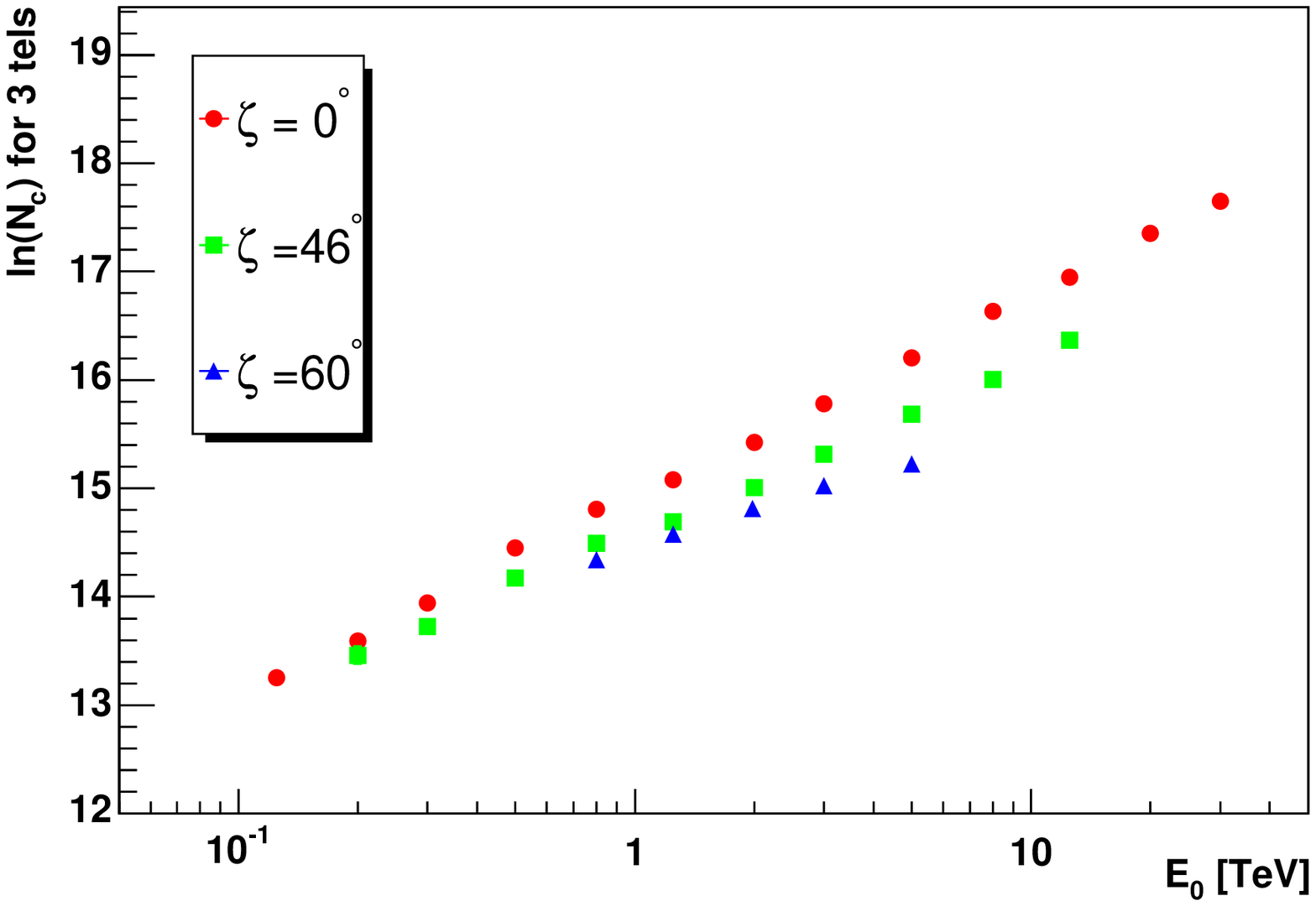,width=\linewidth}
\caption{\it Average value of $\ln N_c$ as a function of $E_0$ for $\gamma$-ray
showers simulated on axis at different zenith angles: $0^{\circ}$ (circles), 
$46^{\circ}$ (squares) and $60^{\circ}$ (triangles), and triggering $\ge$ 3 telescopes.}
\label{fig:calibzen3}
\end{minipage}
\hspace{0.06\linewidth}
\begin{minipage}[t]{0.47\linewidth}
\epsfig{file=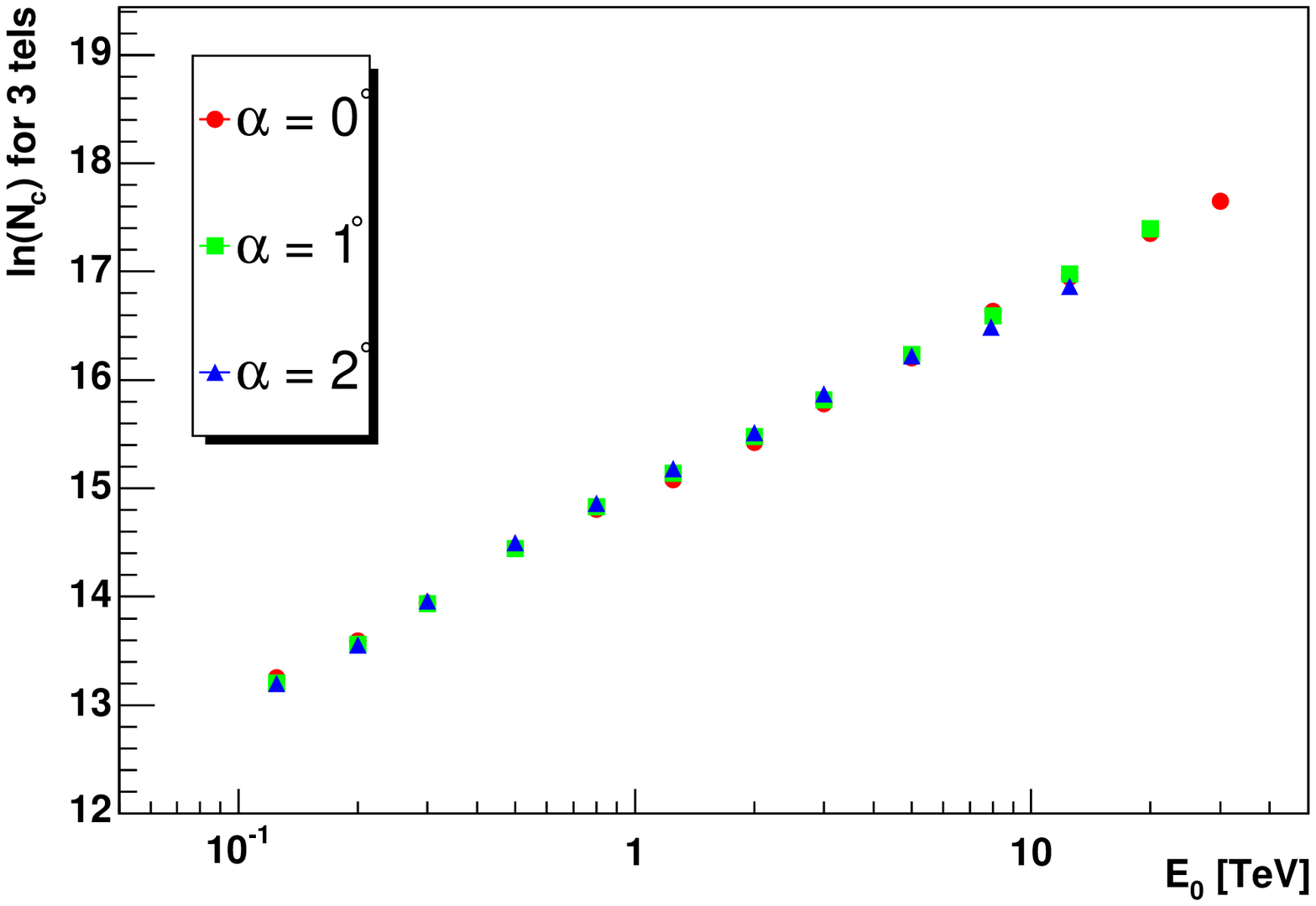,width=\linewidth}
\caption{\it Average value of $\ln N_c$ as a function of $E_0$ for $\gamma$-ray
showers simulated at zenith for two offset angles: $0^{\circ}$ (circles), $1^{\circ}$ (squares)
and $2^{\circ}$ (triangles), and triggering $\ge$ 3 telescopes.}
\label{fig:caliboff3}
\end{minipage}
\end{figure}

However, as an effect of the approximations of the 3D-model used in the fit, 
the coefficients $a$ and $b$ in the formula $\big< \ln N_c \big> = a \ln E_0 +b$ 
depend on the observing conditions; the slight deviation from linearity at the highest energies
for large offset angles is due to shower images not fully contained in the field of view.   
From figures \ref{fig:calibzen} and \ref{fig:caliboff}, one
verifies that the slope $a$ is slightly higher than 1 for showers triggering
2 or 3 telescopes (figures \ref{fig:calibzen2}, 
\ref{fig:caliboff2}, \ref{fig:calibzen3}, \ref{fig:caliboff3}). For a given value of $n_T$,
$a$ is almost independent of the zenith angle and of the offset angle.\\

\begin{figure}
\epsfig{file=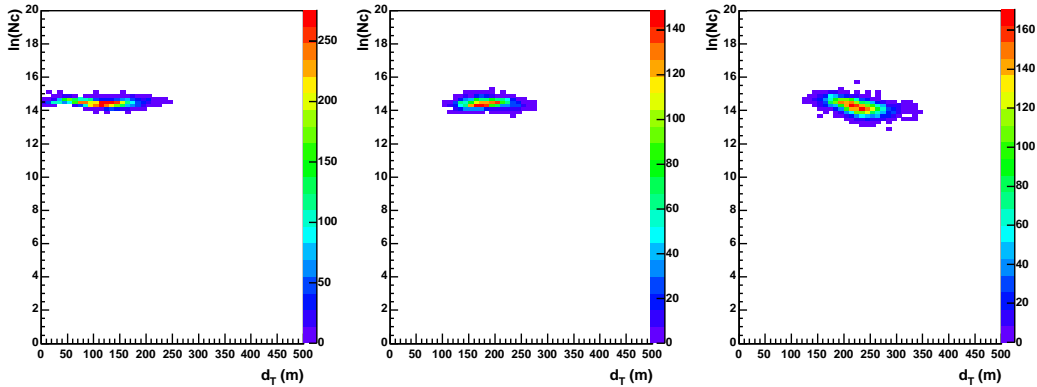,width=\linewidth}
\caption{\it Distribution of $\ln N_c$ versus the impact parameter $d_T$ for 500~GeV $\gamma$-ray showers
simulated on axis at zenith; from left to right: $n_T = 4$, $n_T = 3$, $n_T = 2$.}
\label{fig:dcore500}
\end{figure}

The impact parameter $d_T$, i.e. the distance between the
centre of the array and the shower axis, also affects the sampling of the shower.
In figures \ref{fig:calibzen} to \ref{fig:caliboff3}, $\ln N_c$ was averaged over all 
values of $d_T$ compatible with the trigger conditions; a better energy estimate is 
obtained if the measured value of $d_T$ is taken into account.
For fixed values of $n_T$, $\zeta$ and $\alpha$, the distributions of $\ln N_c$ 
are plotted for different values of $d_T$ in figures \ref{fig:dcore500} (500~GeV
on-axis showers at zenith); 
in most cases, $\big< \ln N_c \big>$ varies only slowly with $d_T$. However, at large distance 
this variation can be more important as one can see on the distribution obtained for showers 
triggering only 2 telescopes, as shown
in figure \ref{fig:dcore500}.\\

\begin{figure}
\epsfig{file=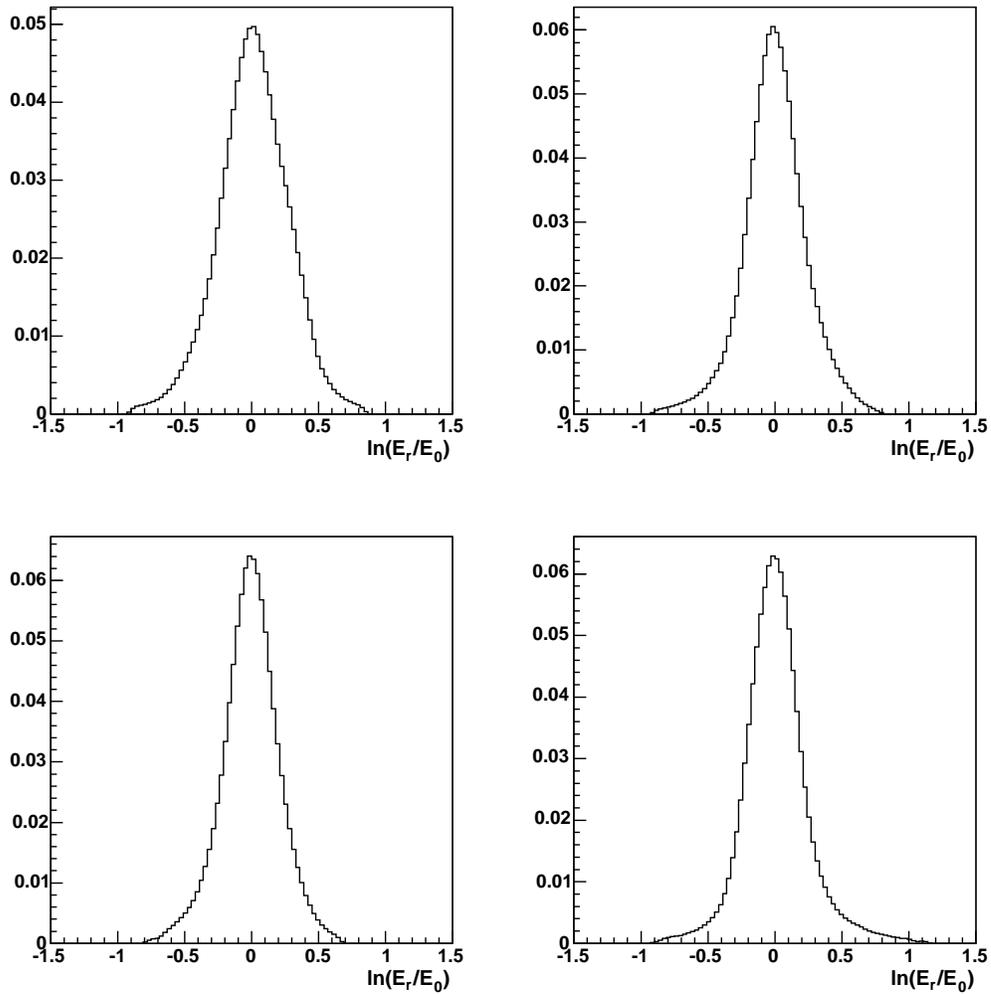,width=\linewidth}
\caption{\it Distribution of $\ln(E_r/E_0)$ for $\gamma$-ray showers simulated on-axis at zenith for different
primary energies: 200~GeV (top left), 500~GeV (top right), 1~TeV (bottom left), 10~TeV (bottom right)}
\label{fig:endis}
\end{figure}

\begin{figure}
\epsfig{file=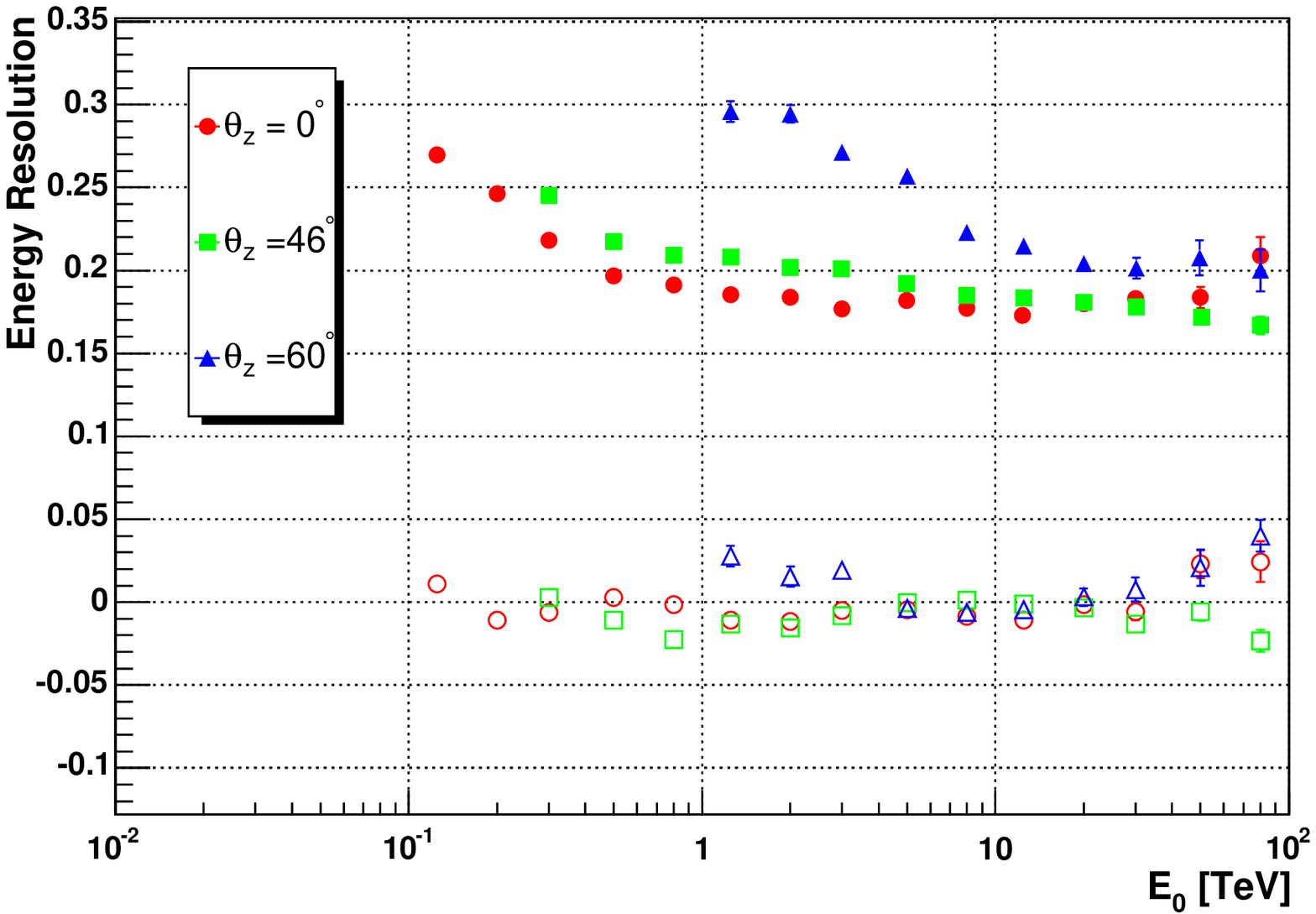,width=\linewidth}
\caption{\it Energy measurement at different zenith angles $\zeta$ and on-axis showers.
Open symbols: bias $\delta$ in $\ln (E_r/E_0)$ as a function of the true primary energy $E_0$; filled symbols:
standard deviation of $\ln (E_r/E_0)$ as a function of $E_0$.Circles are for $\zeta = 0^{\circ}$, squares
for $\zeta = 46^{\circ}$ and triangles for $\zeta = 60^{\circ}$.}
\label{fig:eresolzen}
\end{figure}

\begin{figure}
\epsfig{file=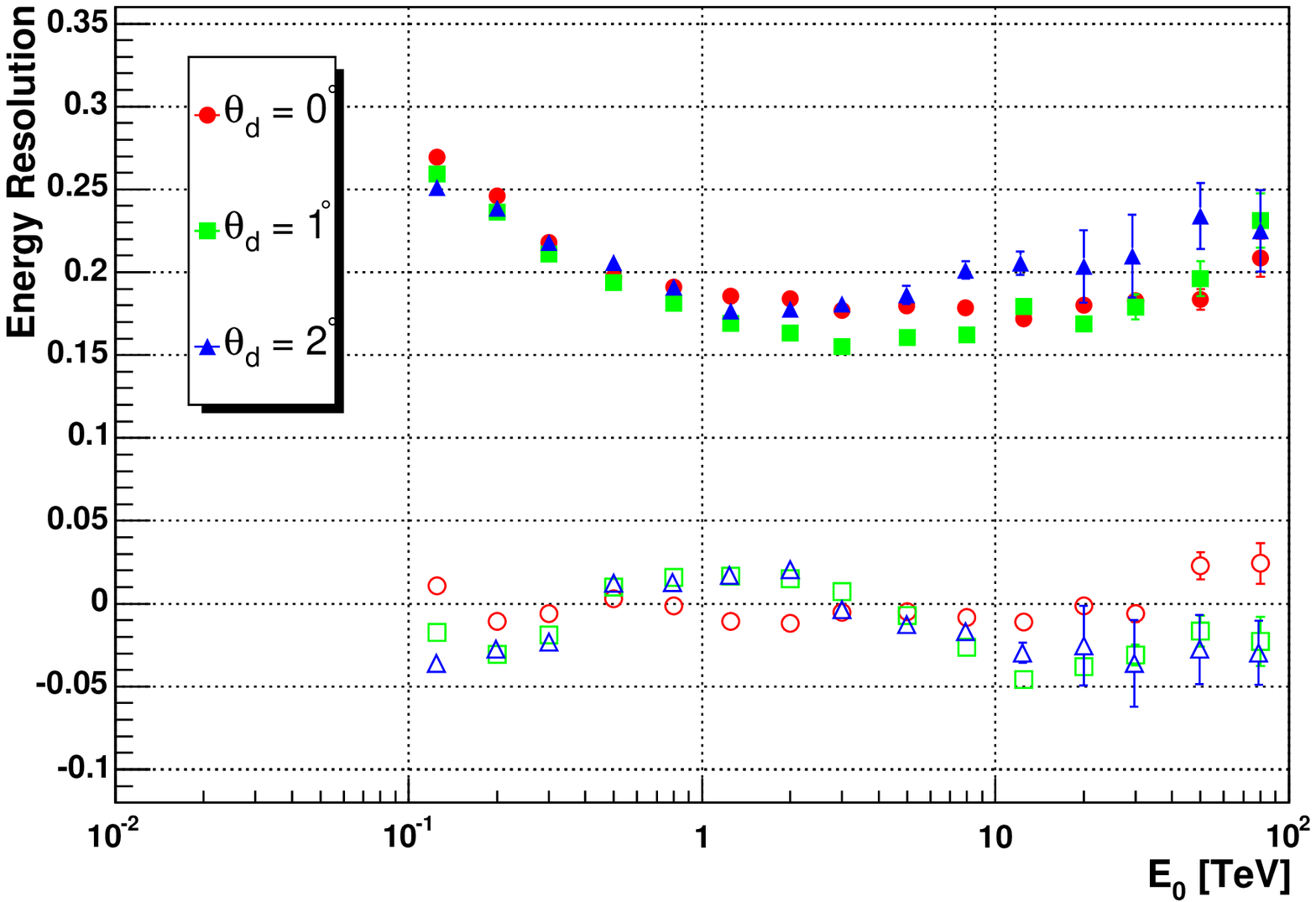,width=\linewidth}
\caption{\it Energy measurement at different offset angles $\alpha$ and showers at zenith.
Open symbols: bias $\delta$ in $\ln (E_r/E_0)$ as a function of the true primary energy $E_0$; filled symbols:
standard deviation of $\ln (E_r/E_0)$ as a function of $E_0$. Circles are for $\alpha = 0^{\circ}$, squares
for $\alpha = 1^{\circ}$ and triangles for $\alpha = 2^{\circ}$.}
\label{fig:eresoloff}
\end{figure}

The energy estimator is based essentially on the value of $N_c$, but also 
on $n_T$, $\zeta$, $\alpha$ and $d_T$. Simulations for fixed
values of these parameters provide a relation between $\big< \ln N_c \big>$ and 
$\ln E_0$ which can be generalized to 
arbitrary values of $\zeta$, $\alpha$ and $d_T$ through interpolations. 
This same relation is then used for a given event satisfying the likelihood fit
to find the energy estimator $E_r$ as a function of $N_c$,
$n_T$, $\zeta$, $\alpha$ and $d_T$. The distribution of
$\ln(E_r/E_0)$ is found to be Gaussian to a good approximation for all observing 
conditions as shown in figure \ref{fig:endis}.
These Gaussian distributions are characterized by their bias $\delta = \big< \ln (E_r/E_0) \big>$ and their
standard deviation $\sigma(\ln (E_r/E_0)) \approx \Delta E_r/E_r$. The variations 
of $\delta$ and of $\Delta E_r/E_r$ with 
energy at different zenith angles are shown in figure \ref{fig:eresolzen} for on-axis showers;
figure \ref{fig:eresoloff} shows the corresponding variations 
at different offset angles for showers at zenith (restricting to $\gamma$-ray showers with
$\theta < 0.1^{\circ}$ and an angular distance from the centre of the camera 
lower than $2^{\circ}$). The bias is smaller than 5$\%$ 
and $\Delta E_r/E_r$ is of the order of 15 to 20\% at low zenith angles.

\section{Conclusion}
The reconstruction method described above differs from the more traditional
ones (ref. \cite{wystan} and \cite{de Naurois03}) in several aspects:
\begin{itemize}
\item The analysis is based on shower parameters in 3 dimensions, not on image
parameters; in this way correlations between different stereoscopic views of the
same shower are taken into account.
\item Gamma-ray candidates are selected on the basis of a few criteria based
on physical properties: rotational symmetry of electromagnetic shower, depth of
shower maximum, lateral spread of the Cherenkov photosphere 
at shower maximum.
\item The distribution of the ratio of the last two variables, namely the 
``reduced 3D-width'' $\omega$, is found to be almost independent of the 
$\gamma$-ray energy and zenith angle and $\gamma$-rays can be efficiently 
selected by requiring $0.8 \times 10^{-3}< \omega < 2 \times 10^{-3}$ 
for all observing conditions.
This criterium provides a $\gamma$-ray/hadron discrimination based on the shower shape
(thus relevant for the study of extended sources) and completely independent of simulations.
\end{itemize}
With a minimal set of natural cuts, the selection efficiency for $\gamma$-rays
as applied to \hess~data is rather uniform between 100~GeV and 10~TeV
and of the order of 80\% (shape criteria) and of 50\% (shape and direction criteria); 
the angular resolution (from 0.04$^{\circ}$ to 0.1$^{\circ}$ at zenith)
and the energy resolution (from 15\% to 20\% in the same conditions)
are comparable to those obtained with the standard \hess~analysis 
\cite{wystan}. This is illustrated in a recent article on the blazar H2356-309~\cite{martin} 
where both analyses were used: using the 3D-model, an excess of 715 $\gamma$-ray candidates is obtained with a significance of $10.9\,\sigma$, to be compared 
with 591 $\gamma$-ray candidates and a significance of $9.7\,\sigma$ with 
the H.E.S.S. standard analysis. Furthermore, requiring at least 3 triggering 
telescopes, the 3D-model yields a higher significance ($11.6\,\sigma$) while 
keeping 453 $\gamma$-ray candidates. 
Thus, the reconstruction method explained above illustrates how the quality
of the stereoscopy, characterized by the telescope multiplicity,
improves the angular resolution and the hadronic rejection of the array, thus
its sensitivity.

\clearpage
\section{Appendix 1: Calculation of the pixel content expected from the 3D-model}
The expected number of Cherenkov photons collected by a given pixel is given by formula 
(\ref{eq:qth}) in which the notations are defined in figure \ref{fig:tele}. The formula
includes the integral $\int_0^{\infty} n_c(r) \, dr$ , taken along the line of sight 
corresponding to the pixel of interest. The calculation of this integral is explicited below.
Let $\vec{p}$ be the unit vector along the line of sight and $\vec{s}$ the unit vector 
along the shower axis, both directed upwards.
Let $\vec{x}_B = \vec{OB}$ be the vector defined by the optical centre O of the telescope 
and the barycentre B of the shower. The parameters $\vec{s}$, $\sigma_L$, $\sigma_T$ and 
$\vec{x}_B$ are provided by the 3D-model. Due to assumption (1) (see section \ref{sec-ass}),
for a shower with $N_c$ Cherenkov photons, the photon density is given by:
\begin{equation}
n_c(r) = \frac{N_c}{(2 \pi)^{3/2} \sigma_L \sigma_T^2} 
\exp \left(- \frac{M}{2}\right)  \:  \: \: \mbox{with} 
\label{eq:density}
\end{equation}
\[ M = \frac{\xi^2}{\sigma_L^2} + \frac{\eta^2}{\sigma_T^2} \: \: \: \mbox{and} \]
\[ \xi = (r \vec{p}-\vec{x}_B) \cdot \vec{s} 
= r \cos \varepsilon - \vec{x}_B \cdot \vec{s} \: \: \: \mbox{and} \]
\[ \eta^2 = (r \vec{p}-\vec{x}_B)^2 - (r \cos \varepsilon - \vec{x}_B \cdot
\vec{s})^2 \: . \]
Defining now the following quantities, independent of $r$:
\[ B_s = \vec{x}_B \cdot \vec{s} \: \: \: ; \: \: \: B_p = \vec{x}_B \cdot \vec{p}
\: \: \: ; \: \: \: u = \cos \varepsilon \: \: \: ; \: \: \: \Delta_B^2 = \vec{x}_B^{\, 2} - B_p^2 \: , \]
\[ \sigma_u^2 = \sigma_T^2 u^2 + \sigma_L^2 (1-u^2) = \sigma_T^2 \cos^2 \varepsilon + \sigma_L^2 
\sin^2 \varepsilon \: \: \: ; \: \: \: \sigma_D^2 = \sigma_L^2 - \sigma_T^2\: , \]
$M$, a second degree polynomial in $r$, can be written in the canonical form:
\[ M  = \frac{\sigma_u^2}{\sigma_L^2 \sigma_T^2} \left[ r - 
\frac{\sigma_L^2 B_p - u B_s \sigma_D^2}{\sigma_u^2} \right]^2 + R \: , \]
$R$ being a quantity independent of $r$ given by:
\[ R =  \frac{1}{\sigma_u^2 \sigma_T^2} \left[ \Delta_B^2 \sigma_u^2 - \sigma_D^2
(uB_p-B_s)^2 \right] \: . \]
Finally, equation (\ref{eq:density}) takes the form:
\[ n_c(r) = \frac{N_c \exp [-R/2) }{(2 \pi)^{3/2} \sigma_L \sigma_T^2} 
\exp \left\{ - \frac{\sigma_u^2}{2 \sigma_L^2 \sigma_T^2} \left[ r - 
\frac{\sigma_L^2 B_p - u B_s \sigma_D^2}{\sigma_u^2} \right]^2 \right\} \: , \]
and the integral in formula (\ref{eq:qth}) is given by:
\begin{equation}
 \int_0^{\infty} n_c(r) \, dr = \frac{N_c C}{2 \pi \sigma_u \sigma_T}
\exp \left\{ - \frac{1}{2} \left[ \frac{\Delta_B^2}{\sigma_T^2} - \frac{\sigma_D^2}
{\sigma_T^2 \sigma_u^2}(uB_p-B_s)^2 \right] \right\} \:  ,
\label{eq:integ}
\end{equation}
in which $C = 1 - {\rm freq} \left( - \, \frac{\sigma_L^2 B_p - \sigma_D^2 u B_s}
{\sigma_u \sigma_T \sigma_L} \right)$, the function ${\rm freq}(x)$ being defined as:
\[ {\rm freq}(x) = \frac{1}{\sqrt{2 \pi}} \int_{-\infty}^x \exp (-t^2/2) \: dt \: \: . \]
\section{Appendix 2: Calculation of the likelihood function}
Since correlations between pixel contents are not taken into account, the likelihood function for
an event takes the form ${\mathcal L} = \prod_i \ell_i$ in which $\ell_i$ is the likelihood function
of the individual pixel $i$. The product is taken over all the selected pixels of the different
images of a given shower. For a given pixel for which the preceding 3D-model predicts an average
number of photoelectron $q_{th}$, the probability to measure a charge (expressed in
number of photoelectrons) in the interval $[q , q+dq]$ is given by:
\[ \sum_{n=0}^{\infty} \frac{\exp(-q_{th}) \, q_{th}^n}{n!} \: \frac{1}{\sqrt{2 \pi} \sigma}
\exp \left( \frac{(q-n)^2}{2 \sigma^2} \right) dq \]
In this formula, Poissonian fluctuations of the effective number $n$ of photoelectrons have been
assumed, as well as Gaussian fluctuations in the phototube response with a standard deviation
$\sigma$. The corresponding probability distribution function at fixed $q$ depends (through
$q_{th}$) on the shower parameters to be fitted and thus represents the factor $\ell$ of the 
likelihood function corresponding to the pixel of interest. In practice, since in 
\hess~ $\sigma = 0.4$~photoelectrons, the sum over $n$ is restricted to the integer values
within a $\pm 3 \sigma$ interval around~$q$.
\section{Acknowledgements}
We thank Pr. W.~Hofmann, spokesman of the H.E.S.S. Collaboration and
Pr. G.~Fontaine, chairman of the Collaboration Board, for allowing us
to use H.E.S.S. data in this publication. We are grateful to Pr. 
T.~Lohse and to Dr. D.~Berge for carefully reading the manuscript and
for providing us with very useful suggestions. 
Finally, our thanks go to all the members of the H.E.S.S. Collaboration 
for their technical support and for many stimulating discussions.  

\end{document}